\documentclass[prd,twocolumn,lengthcheck,showpacs,letterpaper,nofootinbib]{revtex4-1}

\usepackage{color}
\usepackage{graphicx}  
\usepackage{float}
\usepackage{amsmath}
\usepackage{times}
\usepackage{bm}

\newcommand{\Sum}{\displaystyle\sum\limits}

\newcommand{\ii}{{\rm i}}
\newcommand{\D}{\mathrm{d}}
\newcommand{\eff}{\mathrm{eff}}

\newcommand{\real}{\mathrm{real}}
\newcommand{\peak}{\mathrm{peak}}

\newcommand{\RD}{\mathrm{RD}}
\newcommand{\Olap}{\mathcal{O}}
\newcommand{\FF}{\mathcal{FF}}
\newcommand{\MM}{\mathrm{MM}}

\newcommand{\X}{\mathrm{X}}

\def\l({\left(}
\def\r){\right)}

\begin{document}


\title{Template banks to search for low-mass binary black holes in advanced
gravitational-wave detectors.}

\author{Duncan A.\ Brown}
\affiliation{Department of Physics, Syracuse University, Syracuse NY 13244}

\author{Prayush Kumar}
\affiliation{Department of Physics, Syracuse University, Syracuse NY 13244}

\author{Alexander H.\ Nitz}
\affiliation{Department of Physics, Syracuse University, Syracuse NY 13244}

\begin{abstract}
Coalescing binary black holes (BBHs) are among the most likely sources for the
Laser Interferometer Gravitational-wave Observatory (LIGO) and its
international partners Virgo and KAGRA. Optimal searches for BBHs require
accurate waveforms for the signal model and effectual template banks that
cover the mass space of interest. We investigate the ability of the
second-order post-Newtonian TaylorF2 hexagonal template placement metric to
construct an effectual template bank, if the template waveforms used are
effective one body waveforms tuned to numerical relativity (EOBNRv2). We find
that by combining the existing TaylorF2 placement metric with EOBNRv2
waveforms, we can construct an effectual search for BBHs with component masses
in the range $3 M_\odot \le m_1, m_2 \le 25\, M_\odot$.  We also show that the
(computationally less expensive) TaylorF2 post-Newtonian waveforms can be used
in place of EOBNRv2 waveforms when $M \lesssim 11.4\,M_{\odot}$. Finally, we
investigate the effect of modes other than the dominant $l = m = 2$ mode in
BBH searches. We find that for systems with $(m_1/m_2)\leq 1.68$ or inclination
angle: $\iota\leq 0.31$ or $\iota\geq 2.68$ radians, there is no
significant loss in the total possible signal-to-noise ratio due to neglecting
modes other than $l = m = 2$ in the template waveforms. For a source population
uniformly distributed in spacial volume, over the entire sampled region of the
component-mass space, the loss in detection rate (averaged over a uniform distribution
of inclination angle and sky-location/polarization angles), 
remains below $\sim 11\%$. For binaries with high mass-ratios \textit{and} 
$0.31\leq\iota\leq 2.68$, including higher order modes could increase the 
signal-to-noise ratio by as much as $8\%$ in Advanced LIGO. 
Our results can be used to construct matched-filter searches in Advanced LIGO 
and Advanced Virgo.
\end{abstract}

\maketitle

\section{Introduction}
Over the last decade, there has been tremendous progress towards the
first direct detection of gravitational waves. Construction of the Advanced
Laser Interferometer Gravitational-Wave Observatory (aLIGO) is underway, with
completion scheduled for 2014~\citep{Harry:2010zz}. Similar upgrades to
the French-Italian Virgo detector~\citep{aVIRGO} have commenced and construction
of the Japanese KAGRA detector has begun~\citep{Somiya:2011np}.  When these
second-generation gravitational-wave detectors reach design sensitivity, they
will increase the observable volume of the universe by a thousandfold or
more~\citep{aLIGOsensitivity}, compared to the first-generation detectors.

The inspiral and merger of binary black holes (BBHs) are expected to be an important source for detection by
aLIGO~\citep{300yrsofGravitation}.  The rate of BBH coalescences that will be
observed by aLIGO at design sensitivity is estimated to be between
$0.2\,\mathrm{yr}^{-1}$ and $1000\,\mathrm{yr}^{-1}$~\citep{LSCCBCRates2010}.
Accurate knowledge of the gravitational-wave signals generated by BBHs is
crucial for detecting and extracting information about these sources.  
To provide such waveforms, the effective one body (EOB)
model~\citep{EOBOriginalBuonannoDamour} has been calibrated to numerical
simulations of black hole
mergers~\citep{EOBNR01,EOBNRdevel01,EOBNRdevel02,EOBNRdevel03,EOBNRdevel04,EOBdevel01,EOBdevel02,BuonannoEOBv2Main}.
A new EOB waveform family (called EOBNRv2) has been recently proposed that
incorporates information from several non-spinning BBH simulations, with black
hole ring-down quasi-normal modes~\citep{BHRDQNMs,BHPTMinoSasaki} attached to
provide a complete BBH waveform~\citep{BuonannoEOBv2Main}.  The EOBNRv2
waveform is believed to be sufficiently accurate to search for non-spinning
BBH signals in the aLIGO sensitive band (10-1000 Hz). 

Past searches for
BBHs~\citep{Colaboration:2011nz,Abadie:2010yb,Abbott:2009qj,Abbott:2009tt,Messaritaki:2005wv}
used matched-filtering~\citep{Wainstein:1962,Allen:2005fk} to search for
coalescing compact binaries. These searches divided the BBH mass space into a
\emph{low-mass} region with $M = m_1 + m_2 \lesssim 25\, M_\odot$ and a
\emph{high-mass} region with $M \gtrsim 25 \, M_\odot$. In this paper, we
focus attention on BBH systems with component masses between $3 \, M_{\odot}
\lesssim m_1, m_2 \lesssim 25 M_{\odot}$, which encompasses mass distribution
of black hole candidates observed in low-mass X-ray
binaries~\cite{Ozel:2010su}. aLIGO will be able to detect coalescing BBH
systems with component masses $m_1 = m_2 = 25 \, M_{\odot}$ to a maximum
distance of up to $\sim 3.6$~Gpc.  Since we do not know \emph{a priori} the
masses of BBHs that gravitational-wave detectors will observe, searches use a
\textit{bank} of template waveforms which covers the range of BBH component
masses of interest~\citep{Sathyaprakash:1991mt,Balasubramanian:1995bm}.  This
technique is sensitive to the accuracy of the waveform templates that are used
as filters and the algorithm used to place the template
waveforms~\citep{FittingFactorApostolatos}. An accurate template bank is
required as input for matched filter
searches in the Fouier domain~\cite{Allen:2005fk}, as well as newer search algorithms such as the
singular value
decomposition~\citep{Cannon:2010qh}. 

In this paper, we investigate three items of importance to advanced-detector
BBH searches: First, we study the accuracy of template placement algorithms
for BBH searches using EOBNRv2 waveforms. Optimal template placement requires
a metric for creating a grid of waveforms in the desired region of parameter
space~\citep{OwenTemplateSpacing}, however no analytic metric exists for the
EOBNRv2 waveform. In the absence of such a metric, we construct a template
bank using the second-order post-Newtonian hexagonal placement
algorithm~\citep{SathyaBankPlacementTauN,BabaketalBankPlacement,SathyaMetric2PN,Cokelaer:2007kx}.
This metric is used to place template grid points for the aLIGO zero-detuning
high power sensitivity curve~\citep{aLIGONoiseCurve} and we use EOBNRv2
waveforms at these points as search templates.  We find that the existing
algorithm works well for BBHs with component masses in the range $3 M_\odot
\le m_1, m_2 \le 25\, M_\odot$.  For a template bank constructed with a
minimal match of $97\%$ less than $1.5\%$ of non-spinning BBH signals have a
mismatch greater than $3\%$. We therefore conclude that the existing bank
placement algorithm is sufficiently accurate for non-spinning BBH searches in
this mass region.  Second, we investigate the mass range in which the
(computationally less expensive) third-and-a-half-order TaylorF2
post-Newtonian
waveforms~\citep{Sathyaprakash:1991mt,Cutler:1994ys,Droz:1999qx,PNFluxEnergy3PN01,PNFluxEnergy3PN02,Jaranowski:1999qd,Jaranowski:1999ye,Damour:2001bu,KidderPN,Blanchet3PN}
can be used without significant loss in event rate, and where full
inspiral-merger-ringdown EOBNRv2 waveforms are required. We construct a
TaylorF2 template bank designed to lose no more than $3\%$ of the matched
filter signal-to-noise ratio and use the EOBNRv2 model as signal waveforms.
We find that for non-spinning BBHs with $M \lesssim 11.4\,M_{\odot}$, the
TaylorF2 search performs as expected, with a loss of no more than $10\%$ in
the event rate. For higher masses larger event rate losses are observed. A
similar study was performed in Ref.~\citep{CompTemplates2009} using an older
version of the EOB model and our results are quantitatively similar. We
therefore recommend that this limit is used as the boundary between TaylorF2
and EOBNRv2 waveforms in Advanced LIGO searches.  Finally, we investigate the
effect of modes other than the dominant $l = m = 2$ mode on BBH searches in
aLIGO.  The horizion distance of aLIGO (and hence the event rate) is computed
considering only the dominant mode of the emitted gravitational waves, since
current searches only filter for this mode~\citep{LSCCBCRates2010}. However,
the inclusion of sub-dominant modes in gravitational-wave template could
increase the reach of aLIGO~\cite{McKechan:2011ps,Pekowsky:2012sr}. If we
assume that BBH signals are accurately modeled by the EOBNRv2 waveform
including the five leading modes, we find that for systems with $(m_1/m_2)\leq
1.68$ or inclination angle: $\iota \geq 2.68$ or $\iota \leq 0.31$ radians, 
there is no significant loss in the total possible signal-to-noise ratio
due to neglecting modes other than $l = m = 2$ in the template waveforms,
if one uses a $97\%$ minimal-match bank placed using the hexagonal bank placement
algorithm ~\citep{SathyaBankPlacementTauN,BabaketalBankPlacement,SathyaMetric2PN,Cokelaer:2007kx}.
However, for systems with mass-ratio ($q$) $\geq 4$ and $1.08\leq\iota\leq 2.02$, including higher order modes could increase
the signal-to-noise ratio by as much as $8\%$ in aLIGO. This increase in
amplitude may be offset by the increase in false alarm rate from implementing
searches which also include sub-dominant waveform modes in templates, so we encourage the 
investigation of such algorithms in real detector data.

The remainder of this paper is organized as follows: In Sec.~\ref{s:waveforms}
we review the gravitational waveform models used in this study. In
Sec.~\ref{s:results} we present the results of large-scale Monte Carlo signal
injections to test the effectualness of the template banks under
investigation. Finally in Sec.~\ref{s:conclusions} we review our findings and
reccomendations for future work.

\section{Waveforms and Template Bank Placement}
\label{s:waveforms}
\subsection{Waveform Approximants}
The dynamics of a BBH system can be broadly divided into three regimes: (i) The 
early inspiral, when the separation between the black holes is large and their
velocity is small, can be modeled
using results from post-Newtonian (PN)
theory~\citep{PNtheoryLivingReviewBlanchet}. The gravitational-wave phasing of
non-spinning binaries is available up to 3.5PN
order~\citep{PNFluxEnergy3PN01,PNFluxEnergy3PN02,Jaranowski:1999qd,Jaranowski:1999ye,Damour:2001bu,KidderPN,Blanchet3PN}.
(ii) Accurately modeling the late-inspiral and merger requires the numerical
solution of the Einstein
equations~\citep{Pretorius2005,Pretorius2006,BBHNRScheel,BBHNRGonzalezq10,BBHNRPollney,BBHNRLoustoq10,Buchman:2012dw}.
(iii) The final ring-down phase can be modeled using a super-position of
quasi-normal modes (QNMs) which describe the oscillations of the perturbed Kerr
black-hole that is formed from the
coalescence~\citep{BHRDQNMs,BHPTMinoSasaki}. 

Numerical simulation of BBH systems are computationally expensive, and results
are only available for a relatively small number of binary systems (see
e.g.~\citep{Ajith:2012tt}).  The EOB model~\citep{EOBOriginalBuonannoDamour}
provides a framework for computing the gravitational waveforms emitted during
the inspiral and merger of BBH systems.  By attaching a QNM waveform and
calibrating the model to numerical relativity (NR) simulations, the EOB
framework provides for accurate modeling of complete BBH waveforms (EOBNR). The
EOBNR waveforms can be computed at relatively low cost for arbitrary points in the
waveform parameter
space~\citep{EOBNR01,EOBNRdevel01,EOBNRdevel02,EOBNRdevel03,EOBNRdevel04,EOBdevel01,EOBdevel02,BuonannoEOBv2Main}.
In particular the EOB model has recently been tuned against high-accuracy
numerical relativity simulations of non-spinning BBHs of mass-ratios
$q=\{1,2,3,4,6\}$, where $q\,\equiv \, m_1/m_2$~\citep{BuonannoEOBv2Main}; we
refer to this as the EOBNRv2 model, which we review the major features of
below. Throughout, we set $G=c=1$.

The EOB approach maps the fully general-relativistic dynamics of the two-body
system to that of an \textit{effective} mass moving in a deformed
Schwarzschild spacetime~\citep{EOBOriginalBuonannoDamour}. The physical
dynamics is contained in the deformed-spacetime's metric coefficients, the EOB
Hamiltonian~\citep{EOBOriginalBuonannoDamour}, and the radiation-reaction
force. In polar coordinates $(r,\Phi)$, the EOB metric is written as
\begin{equation}\label{eq:dsEOB}
\D s_{\eff}^2 = -A(r)\D t^2 + \dfrac{A(r)}{D(r)}\D r^2 + r^2\left(\D\Theta^2 + \sin^2\Theta \D\Phi^2\right).
\end{equation}
The geodesic dynamics of the \textit{effective} mass $\mu\,=\,m_1 m_2 /
M$ in the background of Eq.~\eqref{eq:dsEOB} is described by an effective
Hamiltonian $H^{\eff}$~\citep{EOBOriginalBuonannoDamour,PadeAD}.
The EOBNRv2 model uses Pade-resummations of the third-order post-Newtonian
Taylor expansions of the metric coefficients $A(r)$ and $D(r)$, with
additional 4PN and 5PN coefficients that are
calibrated~\citep{EOBNRdevel01,EOBNRdevel02,EOBNRdevel03,EOBNRdevel04,BuonannoEOBv2Main} 
to ensure that the dynamics agrees closely with NR simulations of comparable
mass binaries.

Gravitational waves carry energy and angular momentum away from the binary,
and the resulting radiation-reaction force $\hat{F}_{\Phi}$ causes the orbits
to shrink. This is related to the energy flux as 
\begin{equation}
\hat{F}_{\Phi} = -\dfrac{1}{\eta \hat{\Omega}} \dfrac{\D E}{\D t} = -\dfrac{1}{\eta v^3} \dfrac{\D E}{\D t},
\end{equation}
where, $v=(\hat{\Omega})^{1/3}=(\pi Mf)^{1/3}$ and $f$ is the instantaneous
gravitational-wave frequency. The energy flux $\D E/\D t$ is obtained by
summing over the contribution from each term in the multipole expansion of the
waveform, i.e. 
\begin{equation}
\frac{\D E}{\D t} = \frac{\hat{\Omega}^2}{8\pi} \Sum_{l}\Sum_{m} \left|\frac{\mathcal{R}}{M} h_{lm}\right|^2.
\end{equation}
$\mathcal{R}$ is the physical distance to the binary, and $h_{lm}$ are
the multipoles of the waveform when it is decomposed in spin weighted
spherical harmonic basis as
\begin{equation}
h_+ - \ii h_{\times} = \dfrac{M}{\mathcal{R}} \Sum^{\infty}_{l=2} \Sum^{m=l}_{m = -l} Y^{lm}_{-2}\, h_{lm},
\end{equation}
where $Y^{lm}_{-2}$ are the spin weighted spherical harmonics, and $h_+$ and
$h_{\times}$ are the two orthogonal gravitational wave polarizations. These
waveform multipoles depend on the coordinates and their conjugate momenta, and
their Taylor expansions were re-summed as products of individually re-summed
factors~\citep{DamourFluxhlm01}, 
\begin{subequations}\label{eq:hlmdef}
\begin{align}
h_{lm} &= h^F_{lm} N_{lm}\label{eq:hNQC},\\
h^F_{lm} &= h^{(N,\epsilon)}_{lm} \hat{S}_{\eff}^{(\epsilon)} T_{lm} e^{\ii\delta_{lm}} (\rho_{lm})^l\label{eq:hNoNQC};
\end{align}
\end{subequations}
where $\epsilon$ is $0$ if $\left( l+m\right)$ is even, and is $1$ otherwise. This
factorized-re-summation of the waveform multipoles ensures agreement with NR
waveform multipoles~\citep{EOBNRdevel01,EOBNRdevel02,EOBNR01}.  The first
factor $h^{(N,\epsilon)}_{lm}$ is the re-summation of the Newtonian order
contribution and the second factor  $\hat{S}_{\eff}^{(\epsilon)}$ is the
source term, given by the mass or the current moments of the binary in the EOB
formalism~\citep{DamourFluxhlm01,BuonannoEOBTerms}. The tail term $T_{lm}$ is
the re-summation of the leading order logarithmic terms that enter into the
transfer function of the near-zone multipolar waves to the
far-zone~\citep{BuonannoEOBTerms}. The last term $N_{lm}$ attempts to capture
the non-circularity of the quasi-circular orbits.  While calculating the
energy flux in this study we follow exactly the prescription of
Ref.~\citep{BuonannoEOBv2Main}, which calibrates the coefficients of the flux
so that resulting EOB waveform multipoles reproduce their NR counterparts with
high accuracy.

We use the EOBNRv2 Hamiltonian and flux in the equations of motion for the
binary, given by
\begin{subequations}\label{eq:EOBHamiltonianEqs}	
 \begin{align}
\dfrac{\D r}{\D\hat{t}} &\equiv \dfrac{\partial \hat{H}^{\real}}{\partial p_r} = \dfrac{A(r)}{\sqrt{D(r)}}\dfrac{\partial \hat{H}^{\real}}{\partial p_{r*}} (r, p_{r*}, p_{\Phi}) ,\\
\dfrac{\D\Phi}{\D\hat{t}} &\equiv \hat{\Omega} = \dfrac{\partial \hat{H}^{\real}}{\partial p_{\Phi}} (r, p_{r*}, p_{\Phi}) , \\ 
\dfrac{\D p_{r_*}}{\D\hat{t}} &= -\dfrac{A(r)}{\sqrt{D(r)}} \dfrac{\partial \hat{H}^{\real}}{\partial r} (r, p_{r*}, p_{\Phi}) ,\\
\dfrac{\D p_{\Phi}}{\D\hat{t}} &= \hat{F}_{\Phi}(r, p_{r*}, p_{\Phi}) ;
  \end{align}
\end{subequations}
where, $\hat{t}\,(\equiv t/M)$ is time in dimensionless units. 

To obtain the initial values of the coordinates $(r,\Phi,p_{r_*},p_{\Phi})$ that the
system starts out in, we use the conditions for motion on spherical orbits derived in
Ref.\cite{Buonanno:2005xu}, where they treat the case of a generic precessing binary.
We take their non-spinning limit to define the initial configuration of the binary, requiring
\begin{subequations}
\begin{align}\label{eq:IniHr}
\dfrac{\partial\hat{H}^{\real}}{\partial r} &= 0,\\ \label{eq:IniHpr}
\dfrac{\partial\hat{H}^{\real}}{\partial p_{r_*}} &= \dfrac{1}{\eta}\dfrac{\D E}{\D t}\dfrac{(\partial^2\hat{H}^{\real}/\partial r\partial p_{\Phi} )}{(\partial\hat{H}^{\real}/\partial p_{\Phi})(\partial^2\hat{H}^{\real}/\partial r^2)}, \\\label{eq:IniHpphi}
\dfrac{\partial\hat{H}^{\real}}{\partial p_{\Phi}} &= \hat{\Omega}_0,
\end{align}
\end{subequations}
where $\hat{\Omega}_0 = \pi Mf_0$, with $f_0$ being the starting gravitational wave frequency. Simplifying Eq.\eqref{eq:IniHr}, and ignoring the terms involving $p_{r_*}$, as $p_{r_*}\ll p_{\Phi}/r$ in the early inspiral, we get a relation between $p_{\Phi}$ and $r$:
\begin{equation}
p_{\Phi}^2 = \dfrac{r^3A'(r)}{2A(r) - rA'(r)} \\ \label{eq:Inipphi},
\end{equation}
where the prime($'$) denotes $\partial/\partial r$. Substituting this in Eq.\eqref{eq:IniHpphi}, we get the relation:
\begin{equation}
\dfrac{A'(r)}{2r\left(1 + 2\eta\left(\dfrac{A(r)}{\sqrt{A(r)-\frac{1}{2}r\,A'(r)}} - 1\right)\right)} = \hat{\Omega}_0^2.\\ \label{eq:Inir}
\end{equation} 
Thus, between Eq.\eqref{eq:Inir} and Eq.\eqref{eq:Inipphi}, we get the initial values 
of $(r, p_{\Phi})$, corresponding to the initial gravitational wave frequency $f_0$,
and by substituting these into Eq.~\ref{eq:IniHpr}, we obtain the initial value 
of $p_{r_*}$.
With these values, we integrate the equations of motion to obtain the evolution 
of the coordinates and momenta $(r(t),\Phi(t),p_r(t),p_{\Phi}(t))$ over the 
course of inspiral, until the light-ring is reached. In the EOB model, the 
light-ring is defined as the local maximum of the orbital frequency 
$\hat{\Omega}$. From the coordinate evolution, we also calculate $h^F_{lm}(t)$,
which is the analytic expression
for the waveform multipole without the non-quasi-circular correction factor
(defined in Eq.~\eqref{eq:hNoNQC}). While generating $h^F_{lm}(t)$ from the
dynamics, the values for the free parameters in the expressions for
$\delta_{lm}$ and $\rho_{lm}$, are taken from Eqn.[38a-19b] of
Ref.~\citep{BuonannoEOBv2Main}, where they optimize these parameters to
minimize the phase and amplitude discrepancy between the respective EOB
waveform multipoles and those extracted from NR simulations.

The EOB ringdown waveform is modeled as a sum of $N$ quasi-normal-modes
(QNMs)~\citep{EOBNRdevel01,EOBNRdevel02,EOBNRdevel04,BHRDQNMs}
\begin{equation}
h_{lm}^{\RD}(t) = \Sum^{N-1}_{n=0}A_{lmn}e^{-\ii\sigma_{lmn}(t-t_{lm}^{\mathrm{match}})},
\end{equation}
where $N=8$ for the model we consider. The matching time
$t_{lm}^{\mathrm{match}}$ is the time at which the inspiral-plunge and the
ringdown waveforms are attached and is chosen to be the time at which the
amplitude of the inspiral-plunge part of $h_{lm}(t)$ peaks $\left(\mathrm{i.e.}\,
t^{lm}_{\peak}\right)$~\citep{EOBNRdevel01,BuonannoEOBv2Main}. The complex
frequencies of the modes  $\sigma_{lmn}$  depend on the mass $M_f$ and spin
$a_f$ of the BH that is formed from the coalescence of the binary.
We use the relations of Ref.~\citep{BuonannoEOBv2Main}, given by 
\begin{subequations}
\begin{align}
\dfrac{M_f}{M} &= 1 + \left(\sqrt{\frac{8}{9}}-1\right)\eta - 0.4333\eta^2 - 0.4392\eta^3,\\
\dfrac{a_f}{M} &= \sqrt{12}\eta - 3.871\eta^2 + 4.028\eta^3.
\end{align}
\end{subequations}
Using the mass and spin of the final BH, the complex frequencies of the QNMs
can be obtained from Ref.~\citep{BHRDQNMs}, where these were calculated using
perturbation theory. The complex amplitudes $A_{lmn}$ are determined by a
hybrid-comb numerical matching procedure described in detail in Sec.II C of
Ref.~\citep{BuonannoEOBv2Main}.

Finally, we combine the inspiral waveform multipole $h_{lm}(t)$ and the
ringdown waveform $h^{\RD}(t)$ to obtain the complete inspiral-merger-ringdown
EOB waveform $h^{\textrm{IMR}}(t)$, 
\begin{equation}
h^{\textrm{IMR}}_{lm}(t) = h_{lm}(t)\Theta(t^{\mathrm{match}}_{lm}-t) + h^{\RD}(t)\Theta(t-t^{\mathrm{match}}_{lm}),
\end{equation}
where $\Theta(x)=1$ for $x\geq 0$, and 0 otherwise. These multipoles are
combined to give the two orthogonal polarizations of the gravitational
waveform, $h_+$ and $h_{\times}$, 
\begin{equation}\label{eq:hpcfromhlm}
h_+ - \ii h_{\times} = \dfrac{M}{\mathcal{R}} \Sum_{l} \Sum_{m} Y^{lm}_{-2}(\iota,\theta_c) h^{\mathrm{IMR}}_{lm},
\end{equation}
where $\iota$ is
the inclination angle that the binary's angular momentum makes with the line
of sight, and $\theta_c$ is a fiduciary phase angle. To ensure the correctness
of our results, we wrote independent code  to implement the EOBNRv2 waveform
based solely on the content of Ref.~\citep{BuonannoEOBv2Main}. We then
validated our code against the EOBNRv2 waveform algorithm in the LSC Algorithm
Library (LAL)~\citep{lal}.  
We find agreement
between these two implementations, giving us confidence in both our results and the
correctness of the LAL EOBNRv2 code.

Previous searches for stellar-mass BBHs  with total mass $M \lesssim 25
M_\odot$ in LIGO and Virgo used the restricted TaylorF2 PN
waveforms~\citep{Sathyaprakash:1991mt,Cutler:1994ys,Droz:1999qx}. Since this
waveform is analytically generated in the frequency domain, it has two
computational advantages over the EOBNRv2 model: First, the TaylorF2 model does not require
either the numerical solution of coupled ODEs or a Fourier transform to
generate the frequency domain signal requred by a matched filter. 
We compared the speed of generating and Fourier
transforming EOBNRv2 waveforms, to the speed of generating Taylor F2
waveforms in the frequency domain, and found that the former can be
$\mathcal{O}(10^2)$ times slower than the latter. Second, the TaylorF2 model can be implmented trivially as a
kernel on Graphics Processing Units, allowing search pipelines to leverage
significant speed increases due to the fast floating-point performance of
GPU hardware. We found the generation of TaylorF2 waveforms using GPUs to be
$\mathcal{O}(10^4)$ times faster than generating and Fourier
transforming EOBNRv2 waveforms on CPUs. However, use of the TaylorF2 waveform 
may result in a loss in search efficiency due to inaccuracies of the PN 
approximation for BBHs. To investigate the loss in search efficiency versus 
computational efficiency, we use the restricted TaylorF2 waveform described below.

The Fourier transform of a gravitational waveform $h(t)$
is defined by
\begin{equation}
\tilde{h}(f) = \int^{\infty}_{-\infty}e^{-2\pi \ii ft}h(t)\D t.
\end{equation}
Using the stationary phase approximation~\citep{MatthewsWalker}, the Taylor F2
waveform $\tilde{h}(f)$ can be written  directly in the frequency domain as
\begin{equation}\label{eq:hfSPA}
\tilde{h}(f) = Af^{-7/6}e^{ \ii \Psi(f)},
\end{equation}
where we have kept only the leading-order amplitude terms; this is known as
the restricted PN waveform. The amplitude $A\propto \mathcal{M}_c^{5/6}/\mathcal{R}$, 
where $\mathcal{M}_c$ is the \textit{chirp-mass} of the binary, 
$\mathcal{M}_c = (m_1+m_2)\,\eta^{3/5}$, $\eta=m_1m_2/(m_1+m_2)^2$ is the symmetric mass ratio, and $\mathcal{R}$ is the distance to
the binary. The Fourier phase of the waveform at 3.5PN order is given by~\citep{Sathyaprakash:1991mt,Cutler:1994ys,GW2PN,Blanchet:2001ax,Blanchet:2004ek,Poisson:1995ef,Allen:2005fk}
\begin{equation}
\begin{split}\label{eq:PsiSPA}
\Psi(f)=&2\pi ft_c-\phi_c-\dfrac{\pi}{4} + \dfrac{3}{128}\dfrac{1}{\eta}v^{-5}\left[1 + \left(\dfrac{3715}{756} +\dfrac{55}{9}\eta\right)v^2\right.\\
-&\left. 16\pi v^3+\left(\dfrac{15293365}{508032}+\dfrac{27145}{504}\eta +\dfrac{3085}{72}\eta^2 \right)v^4\right.\\
+&\left.\left(\dfrac{38645}{756}-\dfrac{65}{9}\eta\right)\left(1+3\textrm{log}\left(\dfrac{v}{v_{\textrm{lso}}}\right)\right)\pi v^5\right.\\
+&\left.\left[\dfrac{11583231236531}{4694215680}-\dfrac{640}{3}\pi^2 -\dfrac{6848}{21}\gamma_E\right.\right.\\
-&\left.\left. \dfrac{6828}{21}\textrm{log}(4v)+\left(-\dfrac{15737765635}{3048192}+\dfrac{2255}{12}\pi^2 \right)\eta\right.\right.\\
+&\left.\left.\dfrac{76055}{1728}\eta^2 -\dfrac{127825}{1296}\eta^3\right] v^6\right.\\
+&\left.\left(\dfrac{77096675}{254016}+\dfrac{378515}{1512}\eta -\dfrac{74045}{756}\eta^2 \right)\pi v^7\right],
\end{split}
\end{equation}
where  $v=(\pi
Mf)^{1/3}$ is the characteristic velocity of the binary, and $\gamma$ is
Euler's constant.  The initial conditions are set by starting the waveform
from a given gravitational-wave frequency $f=f_{\mathrm{low}}$ and the
waveform is terminated at the frequency of a test particle at the innermost
stable circular orbit (ISCO) of a Schwarzschild black hole $(r = 6M)$.

\subsection{Bank Placement metric}
The frequency weighted overlap between two waveforms $h_1$ and $h_2$, can be
written as
\begin{equation}\label{eq:overlap}
(h_1|h_2) \equiv 2\int^{f_\mathrm{max}}_{f_\mathrm{min}}\dfrac{\tilde{h}_1^*(f)\tilde{h}_2(f) + \tilde{h}_1(f)\tilde{h}_2^*(f)}{S_n(f)}\D f,
\end{equation}
where $S_n(f)$ is the one-sided power spectral density (PSD) of the detector
noise. 
The normalized overlap between the two waveforms is given by
\begin{equation}
(\hat{h}_1|\hat{h}_2) = \dfrac{(h_1|h_2)}{\sqrt{(h_1|h_1)(h_2|h_2)}}.
\end{equation}
In addition to the two mass-parameters of the binary, this normalized overlap
is also sensitive to the relative phase of coalescence $\phi_c$ and to the
difference in the time of coalescence between the two waveforms $h_1$
and $h_2$, $t_c$. These two parameters ($\phi_c$, $t_c$) can be analytically
maximized over to get the maximized overlap $\Olap$
\begin{equation}\label{eq:maxnormolap}
\Olap(h_1,h_2) = \underset{\phi_c,t_c}{\mathrm{max}}\,\left(\hat{h}_1|\hat{h}_2 e^{\ii(2\pi f t_c - \phi_c)}\right),
\end{equation}
which gives a measure of how ``close'' the two waveforms are in the waveform
manifold. The mismatch $M$ between the same two waveforms is written
as, 
\begin{equation}\label{eq:mismatch}
M(h_1,h_2) = 1 - \Olap(h_1,h_2).
\end{equation}
The match (Eq.~\ref{eq:maxnormolap}) can be regarded as an inner-product on
the space of intrinsic template parameters, and thus one can define a 
\textit{metric} on this space \citep{SathyaMetric2PN,OwenTemplateSpacing} (at the point $\theta_1$) as
\begin{equation}\label{eq:metricdef}
 g_{ij}(\theta_1) = -\frac{1}{2}\left.\frac{\partial ^2\Olap\left(h\left(\theta _1\right),h\left(\theta _2\right)\right)}{\partial \theta _1{}^i\partial \theta _2{}^j}\right\vert_{\theta_1^k=\theta^k_2},
\end{equation}
where $\theta_1$ is the set of intrinsic parameters (i.e. $m_1,m_2$ or
some combination) of the binary. Thus the mismatch between waveforms produced by systems with nearly equal mass parameters can be given by
\begin{equation}
 M(h(\theta),h(\theta + \Delta\theta)) \simeq g_{ij}(\theta)\Delta\theta^i\Delta\theta^j.
\end{equation}
For the TaylorF2 approximant, $h(\theta)$ is given by Eq.~(\ref{eq:hfSPA},
~\ref{eq:PsiSPA}), and hence using Eq.~(\ref{eq:overlap},~\ref{eq:maxnormolap}) we 
can get $\Olap(h(\theta_1),h(\theta_2))$ as an analytic function of $\theta_1$ 
and $\theta_2$ (albeit involving an integral over frequency).
This gives a measure of mismatches between neighbouring points in the manifold 
of the mass-parameters, and hence a hexagonal 2D lattice placement can be used 
in the manifold of the mass parameters \citep{SathyaMetric2PN} (and references therein), 
to construct a geometric lattice based template bank~\citep{SathyaBankPlacementTauN,OwenTemplateSpacing,SathyaMetric2PN}. 
 
On the other hand, for the EOBNRv2 approximant, $h(\theta)$ is obtained through 
numerical solutions of the Hamiltonian equations, Eq.(\ref{eq:EOBHamiltonianEqs}). 
In this case, the calculation of the metric would involve derivatives of coordinate
evolution obtained from numerically integrated equations of motion, 
which could introduce numerical instabilities in the metric. So the concept of 
a metric, as in Eq.~(\ref{eq:metricdef})`, cannot be used in a convenient (semi-) 
analytic form for the construction of a bank with the EOBNRv2 approximant.

\section{Results}
\label{s:results}

To assess the effectualness of the template banks constructed here, we compute
the fitting factors~\citep{FittingFactorApostolatos} of the template bank,
defined as follows. If $h^e_a$ is the waveform emitted by a BBH system then the \textit{Fitting
Factor} of a bank of template waveforms (modeled using approximant $\X$) for
this waveform, is defined as the maximum value of maximized normalized
overlaps between $h^e_a$ and all members $h^{\X}_b$ of the bank of template
waveforms \citep{FittingFactorApostolatos}; i.e.
\begin{equation}\label{eq:defFF}
\mathcal{FF}(a,\X) = \underset{b\, \in\, \mathrm{bank}}{\textrm{max}}\,\Olap(h^e_a,h^{\X}_b).
\end{equation}
This quantity simultaneously quantifies the loss in recovered signal-to-noise
ratio (SNR) due to the discreteness of the bank, and the inaccuracy of the
template model. The similarly defined quantity $\MM$ (minimal match)
quantifies the loss in SNR due to only the discreteness of the bank as both
the \textit{exact} and the template waveform is modeled with the same waveform
model, i.e.
\begin{equation}
\MM = \underset{a}{\textrm{min}}\,\underset{b\, \in\, \mathrm{bank}}{\textrm{max}}\,\Olap(h^{\X}_a,h^{\X}_b),
\end{equation}
where $a$ is any point in the space covered by the bank, and $\X$ is the
waveform approximant. For a detection search that aims at less than $10\%\,
(15\%)$ loss in event detection rate due to the discreteness of the bank and
the inaccuracy of the waveform model, we require a bank of template waveforms
that has $\mathcal{FF}$ above $0.965\,
(0.947)$~\citep{WaveformAccuracy2008,WaveformAccuracy2010,CompTemplates2009}.
Throughout, we use the aLIGO zero-detuning high power noise curve as the
PSD for bank placement and overlap calculations, and set $f_\mathrm{min} =
15$~Hz. The waveforms are generated at a sample rate of $8192$~Hz, and we set 
$f_\mathrm{max} = 4096$~Hz, i.e. the Nyquist frequency.

The expectation value of the SNR for a signal, $\rho$, from a source located at
a distance $\mathrm{D}$ 
is proportional to $1/\mathrm{D}$, which comes from the dependence of the 
amplitude on the distance. In other words, the range to which a
souce can be seen by the detector
\begin{equation}
 D_{\mathrm{obs}} = \dfrac{(g,g)}{\rho^*},
\end{equation}
where $g$ is the GW strain produced by the same source at the detector, when 
located at a unit distance from the detector, and $\rho^*$ is the threshold on 
SNR required for
detection (typically taken as $\rho^* = 8$). For non-precessing binaries, for 
which the sky-location ($\theta,\phi$) and polarization angles ($\psi$) do not change over the course of
inspiral, the effective volume in which the same source can be detected is $\propto\,D_{\mathrm{obs}}^3$ \citep{Finn:1992xs}, i.e.
\begin{equation}
 \mathrm{V}_{\mathrm{obs}} = k\,D_{\mathrm{obs}}^3,
\end{equation}
where the proportionalality constant $k$ comes from averaging over various 
possible sky positions of the
binary. The use of discrete template banks, and lack of knowledge of the 
\textit{true} GW signal model, leads to the observed SNR $\rho'$ being lower 
than the optimal SNR $\rho = (h,h)$, i.e.
\begin{equation}
 \rho' = \FF \, \rho,
\end{equation}
where $\FF$ is the fitting-factor of the template bank employed in the search
for the particular system. The observable volume hence goes down as
\begin{equation}
 \mathrm{V}^{\mathrm{eff}}_{\mathrm{obs}} = k\, (\FF\times D_{\mathrm{obs}})^3.
\end{equation}
If we assume that the source population is distributed uniformly in spacial 
volume in the universe, then the ratio
$\mathrm{V}^{\mathrm{eff}}_{\mathrm{obs}}/\mathrm{V}_{\mathrm{obs}}$ also gives
the fraction of systems within the detector's reach that will be seen by the 
matched-filtering search. For a system with given mass-parameters $\theta_1$, the ratio of
the total $\mathrm{V}^{\mathrm{eff}}_{\mathrm{obs}}$ available to it
for different inclinations and sky-locations, to the total
$\mathrm{V}_{\mathrm{obs}}$ available to it for the same samples of angles, 
will give an estimate of the fraction of systems with those masses (marginalized over other
parameters - they being uniformly distributed) that will be seen by the matched-filter
search. This quantity,
\begin{eqnarray}\label{eq:epsilondef}
 \epsilon_{\mathrm{V}} (\theta_1)&=& \dfrac{\Sum_{\mathrm{\theta_2}}\mathrm{V}^{\mathrm{eff}}_{\mathrm{obs}}(\theta_1,\theta_2)}{\Sum_{\mathrm{\theta_2}}\mathrm{V}_{\mathrm{obs}}(\theta_1,\theta_2)},\nonumber \\ 
 &=& \dfrac{\Sum_{\mathrm{\theta_2}}\FF^3(\theta_1,\theta_2)\mathrm{V}_{\mathrm{obs}}(\theta_1,\theta_2)}{\Sum_{\mathrm{\theta_2}}\mathrm{V}_{\mathrm{obs}}(\theta_1,\theta_2)}, 
\end{eqnarray}
where $\theta_2 = \{\iota,\theta,\phi,\psi\}$ are the parameters being
averaged over,
we will refer to as the \textit{volume-weighted fitting-factor}. It
essentially measures the average of the fractional observable volume
loss, weighted by the actual available observable volume, and so simultaneously 
downweights the loss in the observable volume for binary configurations
to which the detector is relativey less sensitive to begin with.
We can give the parameter sets $\theta_1$ and $\theta_2$ different elements than the ones shown here, i.e. $\theta_1 \neq \{m_1,m_2\}, \theta_2 \neq \{\iota,\theta,\phi,\psi\}, \theta_1 \cap \theta_2 = \{m_1,m_2,\iota,\theta,\phi,\psi\}$, in order to obtain more information about another set of parameters $\theta_1'$.

\subsection{EOBNRv2 templates placed using TaylorF2 metric}
\begin{figure}
	\begin{center}
		\scalebox{1}{\includegraphics[width=\columnwidth]{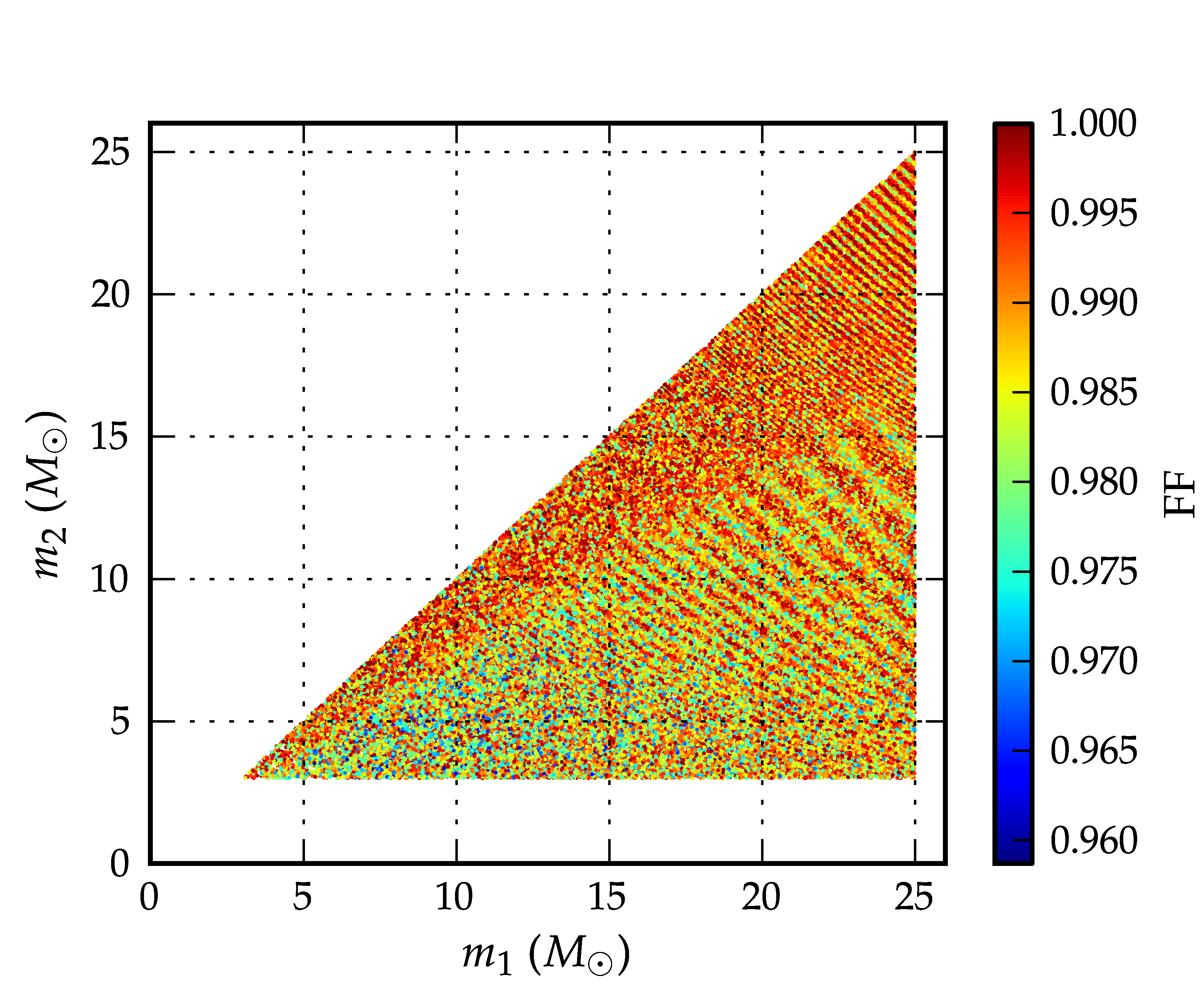}} 
	\end{center}
\caption{This figure shows the effectualness of a bank of EOBNRv2 templates,
placed using the 2PN accurate hexagonal template placement of
Ref.~\citep{BabaketalBankPlacement}, to search for a population of BBH signals
simulated with EOBNRv2 waveforms. The masses of the BBH population are chosen
from a uniform distribution of component masses between $3$ and $25\,
M_{\odot}$. For each injection, we plot the component masses of the injection,
and the fitting factor $\left(\FF\right)$.} \label{fig:match_eobeob_all}
\end{figure}

\begin{figure}
	\begin{center}
		\scalebox{1}{\includegraphics[clip=false, width=\columnwidth]{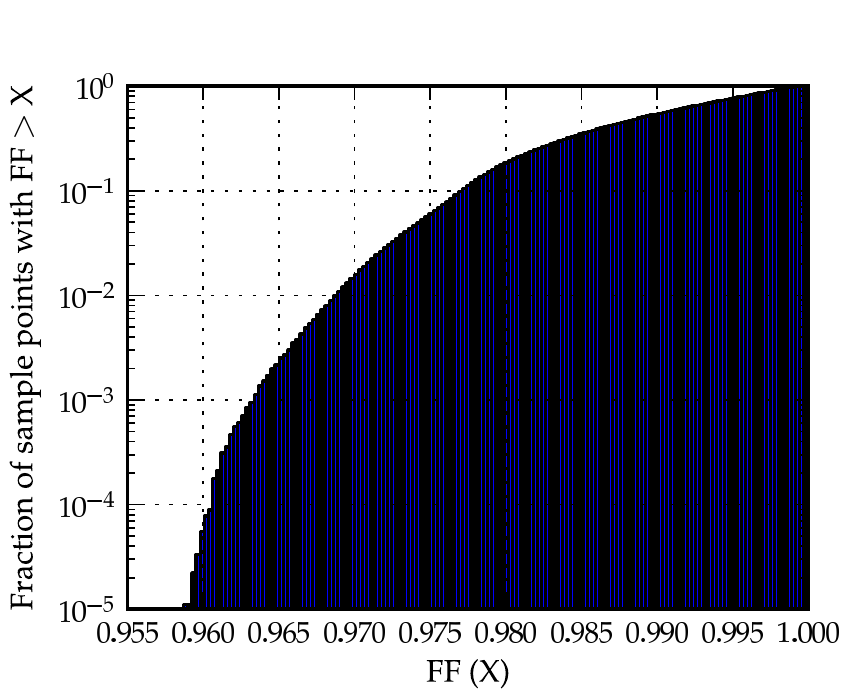}}
	\end{center}
\caption{This figure shows a cumulative histogram of the fraction of the BBH
signal space (on the y-axis), where the bank of EOBNRv2 waveforms has
$\mathcal{FF}$ less than the respective values on the x-axis.  The EOBNRv2
bank has a fitting factor $\mathcal{FF}$ below the $0.97$ for less than $\sim
1.5\%$ of all simulated signals with component-masses $m_1, m_2$ between
$3\, M_\odot$ and $25\,M_{\odot}$.} \label{fig:cumhist_eobeob_all}
\end{figure} 

In this section we measure the effectualness of the second-order
post-Newtonian hexagonal template bank placement metric described in
Ref.~\citep{BabaketalBankPlacement} when used to place EOBNRv2 waveform
templates for aLIGO.  The same template placement algorithm was used to place
a grid of third-and-a-half order post-Newtonian order TaylorF2 waveforms for
low-mass BBH detection searches for initial LIGO and Virgo
observations~\citep{Colaboration:2011nz,Abadie:2010yb,Abbott:2009qj,Abbott:2009tt,Messaritaki:2005wv}.
We construct a template bank which
has a desired minimal match of $0.97$ for waveforms with component masses
between $3 M_\odot \le m_1, m_2 \le 25 M_\odot$. This template
bank contains $10,753$ template grid points in $(m_1,m_2)$ space for the aLIGO noise curve, 
compared to $373$ grid points for the initial LIGO design noise
curve. For the template waveforms at each grid point, we use the EOBNRv2
waveforms, rather than TaylorF2 waveforms.  Since the metric itself was
derived using second-order TaylorF2 waveforms, we do not, \textit{a priori}
know if this metric is a good measure to use to place template banks for
EOBNRv2 waveforms. 

To test the effectualness of this template bank, we perform a Monte-Carlo
simulation over the $3 M_\odot \le m_1, m_2 \le 25 M_\odot$ BBH mass space to
find regions where the bank placement algorithm leads to under-coverage.  We
sample 90,000 points uniformly distributed in individual component masses. For
each of these points, we generate an EOBNRv2 waveform for the system with
component masses given by the coordinates of the point.  We record the $\mathcal{FF}$ of the template
bank for each of the randomly generated BBH waveforms in the Monte-Carlo
simulation.  Since we use EOBNRv2 waveforms both to model the true BBH signals and
as matched-filter templates, any departure in fitting factor from unity is
due to the placement of the template bank grid. 

For a bank of template waveforms constructed with a $\MM$ of $0.97$,
Fig.~\ref{fig:match_eobeob_all} and Fig.~\ref{fig:cumhist_eobeob_all} show
that the $\mathcal{FF}$ of the bank remains above $0.97$ for $\sim 98.5\%$ of
all simulated BBH signals. Less than $\sim 1.5\%$ of signals have a minimal
match of less than 0.97, with the smallest value over the 90,000 sampled
points being $\sim 0.96$.  The diagonal features observed in
Fig.~\ref{fig:match_eobeob_all} are due to the hexagonal bank placement
algorithm and are related to the ellipses of constant chirp mass in Fig.~4  of
Ref.~\citep{BabaketalBankPlacement}.  From these results, we conclude that the
existing template bank placement metric adequately covers the BBH mass space
with EOBNRv2 waveform templates; it is not necessary to construct a metric
specific to the EOBNRv2 model. aLIGO detection searches can employ the
second-order post-Newtonian bank placement metric with the hexagonal placement 
algorithms~\citep{SathyaMetric2PN,SathyaBankPlacementTauN,BabaketalBankPlacement,OwenTemplateSpacing,Cokelaer:2007kx}
to place template banks for EOBNRv2
waveforms without a significant drop in the recovered signal-to-noise ratio.

\subsection{Effectualness of TaylorF2 templates}\label{s2:eob22f2}
\begin{figure}
  \begin{center}
   \scalebox{1}{\includegraphics[width=\columnwidth]{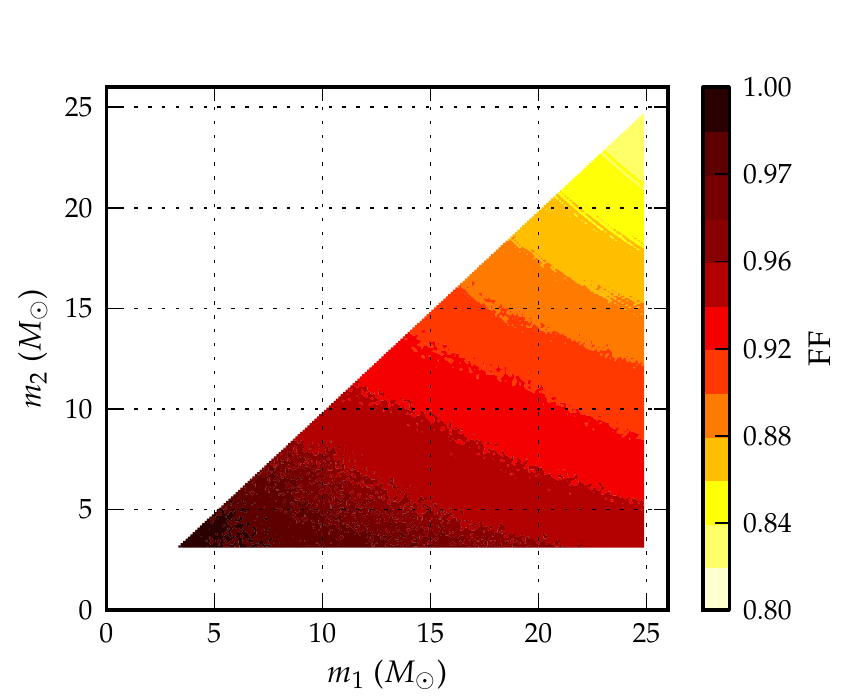}} 
  \end{center}
\caption{\label{fig:match_f2eob_f2eob22_all}The fitting factor $\mathcal{FF}$
of a bank of TaylorF2 waveforms, constructed with $\MM = 0.97$, 
for a population of BBH systems which are modeled using EOBNRv2 signals.} 
\end{figure}

\begin{figure}
\centering
\includegraphics[scale=0.04, clip=false, keepaspectratio=true, width=\columnwidth]{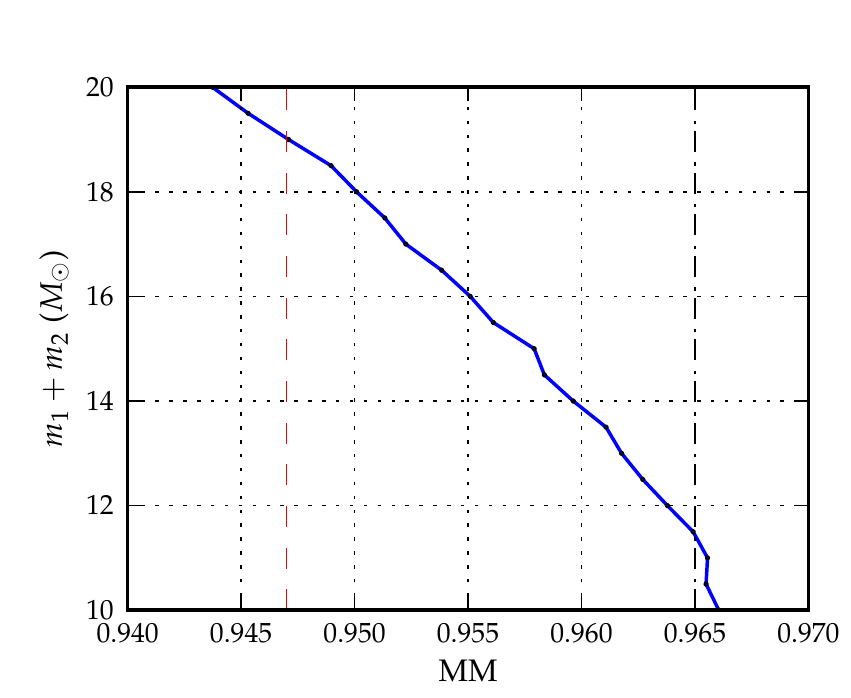} 

\caption{\label{fig:mmvsM_f2eob22}The blue curve shows the upper-bound 
on total-mass for the sub-region over which the TaylorF2 bank has a 
minimal-fitting-factor as given on the x-axis. We observe that the TaylorF2 
bank has a minimal-fitting-factor of $0.965\,(0.947)$ for the region with total
masses below $\sim 11.4 M_{\odot}\,(19 M_{\odot})$. The minimal-fitting-factor
is the fitting-factor value which is less than the fitting-factors of the 
TaylorF2 bank for $\geq 99.75\%$ of the points sampled in the sub-region.}
\end{figure}

We next explore the efficiency of using the computationally cheaper TaylorF2
waveforms to search for a population of BBH signals with
component masses between $(3$--$25) M_\odot$. The signals from this population
are modeled 
with the full EOBNRv2 waveforms. We use the same template bank placement as above,
however now we use the third-and-a-half PN order TaylorF2 model as the
template waveforms. This model does not capture the merger and ringdown of BBH
signals, as it is terminated at the Schwarzchild test-particle ISCO.
Furthermore, it diverges from the true BBH signal in the late inspiral.
It is important to determine when these effects become important.

We sample the $(3$--$25) M_\odot$ BBH component mass space at 100,000 points by
generating an EOBNRv2 waveform to generate the ``true'' signal waveform.  We
generate a bank of TaylorF2 template waveforms over the same region, and
calculate its $\FF$ for each of the sample points, against the corresponding
EOBNRv2 waveform. Fig.~\ref{fig:match_f2eob_f2eob22_all}  shows the
distribution of the $\FF$ obtained for the TaylorF2 bank. Clearly the TaylorF2
bank is not effectual for the enture BBH region considered, with mismatches of
up to $18\%$ observed. We divide the sampled component mass space into
sub-regions which consist of systems with total masses below different thresholds,
and compute the minimal-fitting-factor of the bank over those. In Fig.~\ref{fig:mmvsM_f2eob22}, 
the blue (solid) curve shows the upper-limit on 
total mass for different sub-regions against the minimal-fitting-factor of the 
TaylorF2 bank over those. The minimal-fitting-factor over a sub-region is 
taken to be the fitting-factor value which is less than the fitting-factors
for $\geq 99.75\%$ of the points sampled in the sub-region. We find that the 
TaylorF2 template bank has $\FF$ above $0.965$ ($0.947$) for the region 
with total masses below $11.4M_{\odot}$ ($19M_{\odot}$). We conclude 
that the TaylorF2 bank is effectual for BBH signals below $\sim 11.4M_{\odot}$. 

The value of our limit on total-mass is in agreement with the previous study in
Ref.~\citep{CompTemplates2009}, however this analysis used the EOBNRv1
model~\citep{Buonanno:2007pf} and an older version of the Advanced LIGO noise
curve \citep{CompTemplates2009}. This agreement provides
confidence that this limit will be robust in aLIGO searches and we propose
this limit as the upper cutoff for the computationally cheaper TaylorF2
search. To investigate the loss in the $\FF$ due to the mismatch in the
template and signal waveform models, we also performed a Monte-Carlo
simulation using a denser TaylorF2 bank with $\MM\,=\,0.99$. We found that
using this dense bank of third-and-a-half order TaylorF2 waveforms, we can
relax the limit on the upper mass to $\sim 16.3M_{\odot}$ ($21.8M_{\odot}$) and
still achieve a  $\FF$ above $0.965$ ($0.947$), for over $99.75\%$ of the signals
sampled in the region.  However, 
increasing the minimal match increases the size of the template bank from 
$10,753$ to $29,588$ templates. This is a significant increase, compared to 
the cost of filtering with EOBNRv2 templates.

\subsection{Effect of sub-dominant modes}
\begin{figure*}
\centering
\includegraphics[scale=0.04, clip=false, totalheight=0.3\textheight, width=\columnwidth]{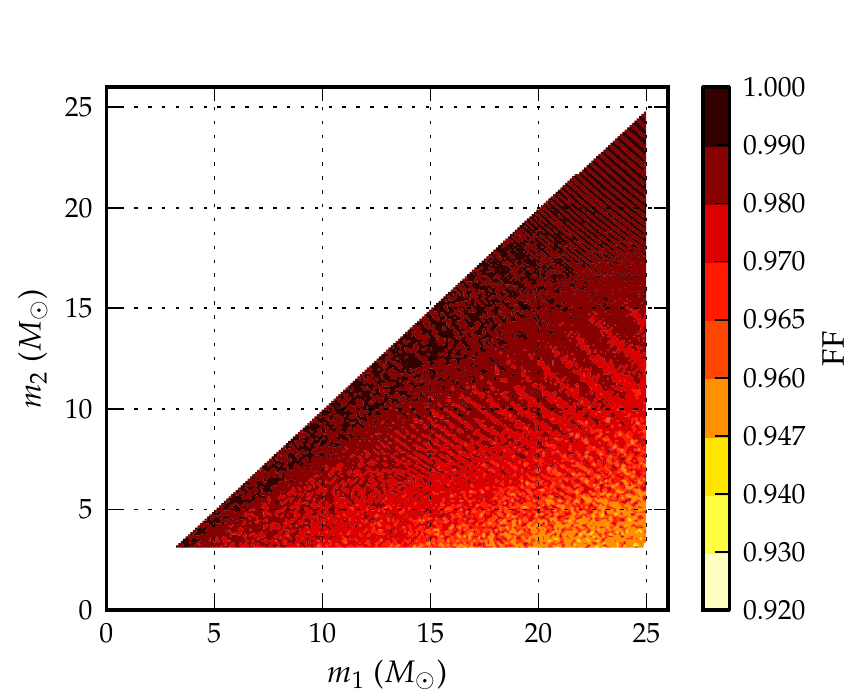} 
\includegraphics[scale=0.04, clip=false, totalheight=0.3\textheight, width=\columnwidth]{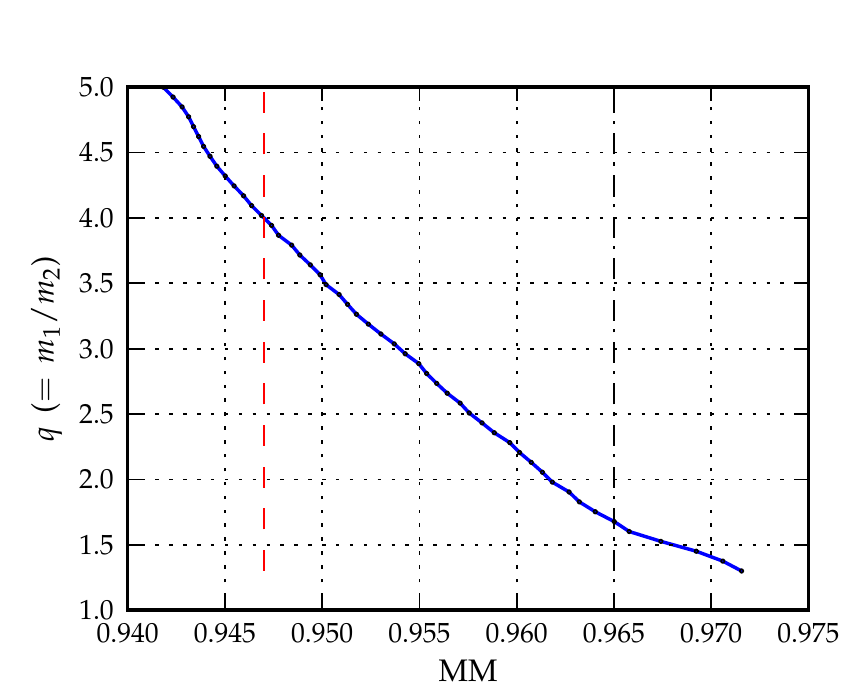} 
\caption{\label{fig:match_eob22eobhm_m1m2} (left) The $\FF$ of a
bank of EOBNRv2 waveforms, constructed with a minimal match of 0.97 at each
point in the stellar-mass BBH component-mass region. While the templates are
modeled as the dominant-mode $l=m=2$ EOBNRv2 waveforms, the signals are modeled 
including the sub-dominant waveform modes as well (EOBNRv2HM). (right) This figure shows the 
upper-bound on mass-ratio ($q$) for the region where a bank of EOBNRv2
templates has a minimal-fitting-factor as given on the x-axis. We observe that for
the region with $q \leq 1.68\,(4)$, the minimum-match of the bank is below $0.965\,(0.947)$. 
From both the figures, we notice a systematic fall in the coverage of the EOBNRv2 
template bank with increasing mass-ratio.} 
\end{figure*}
\begin{figure*}
\centering
\includegraphics[scale=0.04, clip=false, totalheight=0.3\textheight, width=\columnwidth]{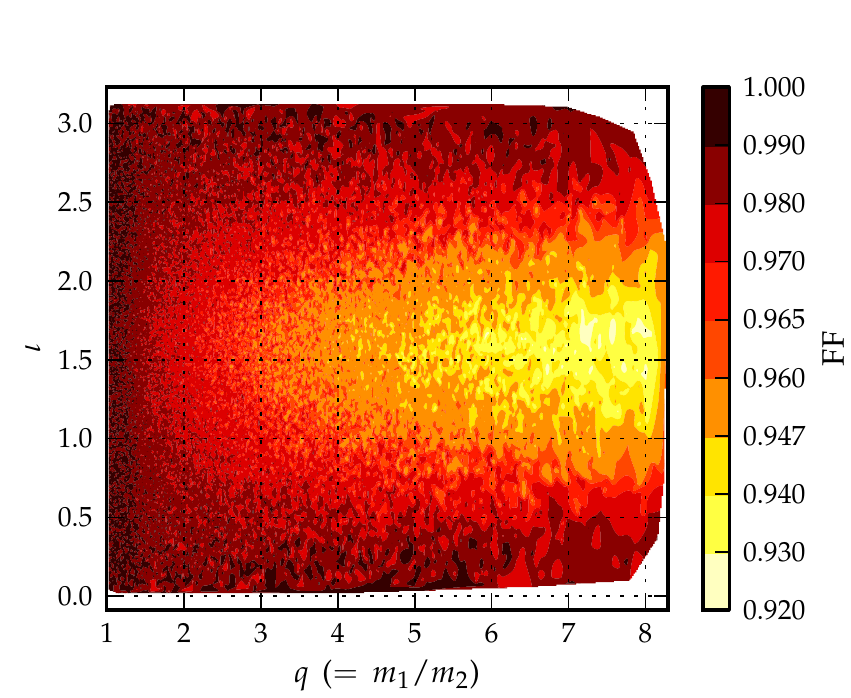} 
\includegraphics[scale=0.04, clip=false, totalheight=0.3\textheight, width=\columnwidth]{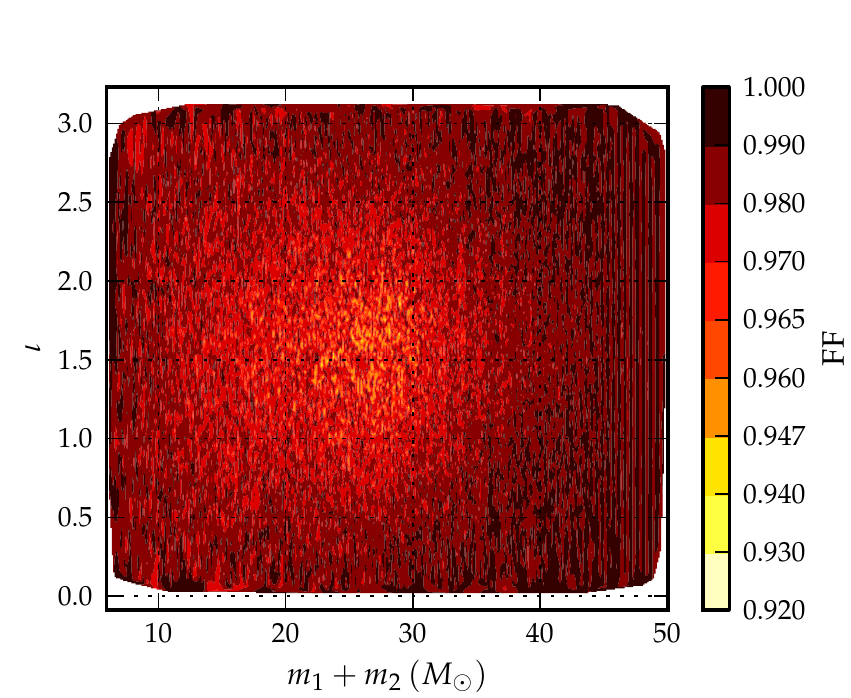} 
\caption{\label{fig:incvsM_eob22eobhm} (left) The $\FF$ of a
bank of EOBNRv2 waveforms, constructed with a minimal match of 0.97 at each
point in the stellar-mass BBH $q-\iota$ space. While the templates are
modeled as the dominant-mode $l=m=2$ EOBNRv2 waveforms, the signals are modeled 
including the sub-dominant waveform modes as well (EOBNRv2HM). We observe a loss in fitting-factors, upto $\sim 8\%$, for 
systems with high mass-ratios ($q$) \textit{and} inclination angle ($\iota$) close to $\pi/2$.
(right) The $\FF$ for the same population of signals, now shown on the $M-\iota$ plane. We observe the loss in fitting factors to be relatively lesser
for more massive binaries.}
\end{figure*}
\begin{figure*}
\centering
\includegraphics[scale=0.04, clip=false, totalheight=0.3\textheight, width=\columnwidth]{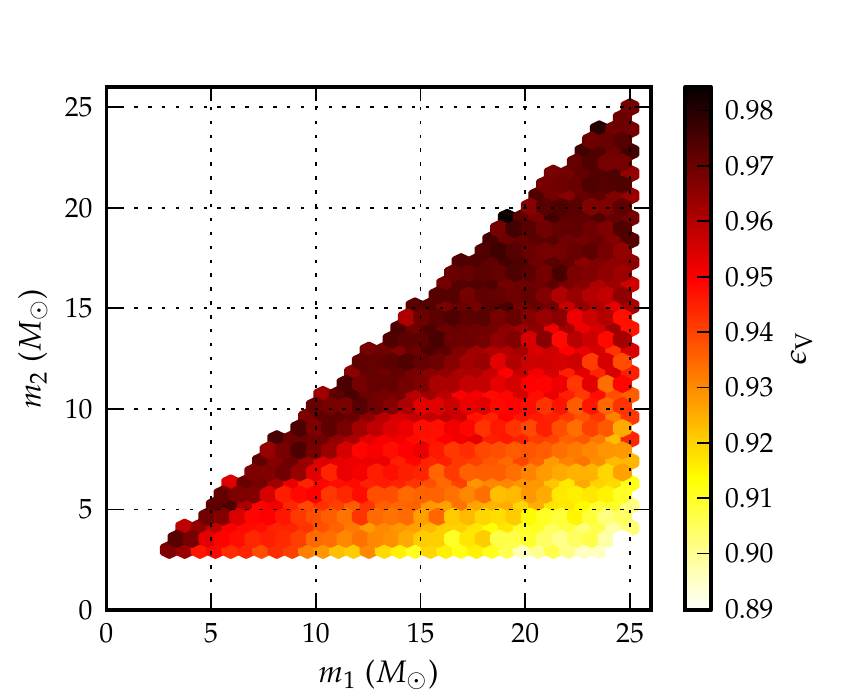} 
\includegraphics[scale=0.04, clip=false, totalheight=0.3\textheight, width=\columnwidth]{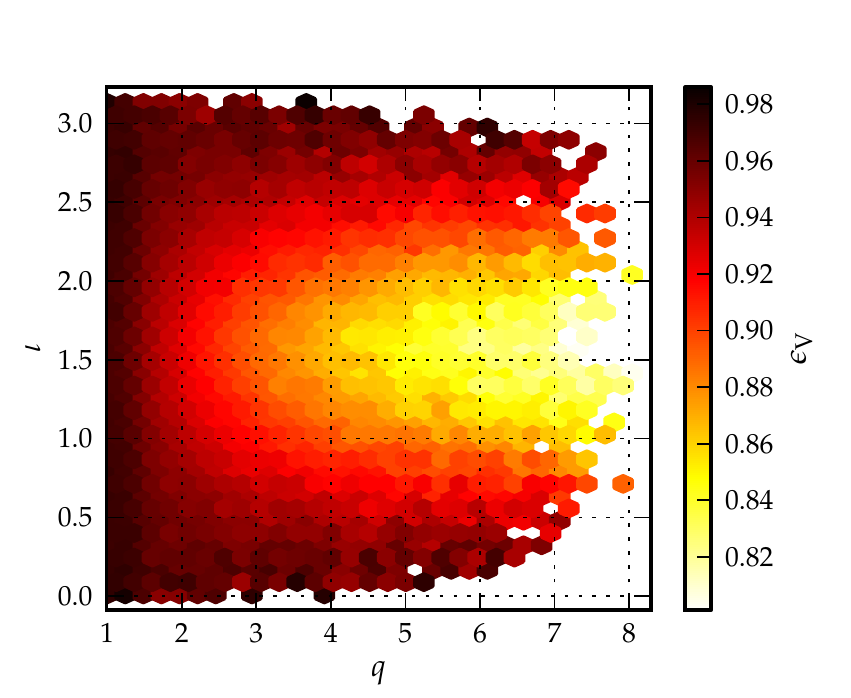} 
\caption{\label{fig:VeffLoss_eob22eobhm} (left) This figure shows $\epsilon_{\mathrm{V}}\left(\theta_1=\{m_1,m_2\}\right)$ in the component-mass
space (see Eq.~\ref{eq:epsilondef}). 
This gives the 
fraction of total observable volume that is visible to a search which uses the
leading order $l=m=2$ EOBNRv2 waveform template bank, placed with the 2PN 
accurate TaylorF2 bank placement metric. For a population of signals, that is distributed uniformly in spacial volume,
this is equivalent to the fraction of the maximum possible event observation rate that we get with the use of a discrete bank of matched-filters. We observe that the loss in event observation rate,
averaged over all parameters (uniformly distributed) but $\theta_1=\{m_1,m_2\}$,
does not exceed $\sim 11\%$ for any region of the component-mass space.
(right) This figure shows $\epsilon_{\mathrm{V}}\left(\theta_1=\{q,\iota\}\right)$ over the $q-\iota$ plane. We note that the 
maximum averaged loss in the detection rate is for systems
with high mass ratios \textit{and} $\iota\in[1.08,2.02]$, and 
can go as high as $\sim 20\%$ for such systems.
} 
\end{figure*}

Having established that the second-order post-Newtonian hexagonal template
bank is effectual for placing a bank of EOBNRv2 templates, we now investigate
the effect of neglecting sub-dominant modes in BBH searches. The sensitivity
reach of the aLIGO detectors is normally computed assuming that the search is
only sensitive to the dominant $l=m=2$ mode of the gravitational waveform. For
binary black hole signals, sub-dominant modes may contain significant
power~\cite{Pekowsky:2012sr}. A search that includes these modes could, in
principle, have an increased reach (and hence event rate) compared to a search
that only uses the dominant mode. The EOBNRv2 model of
Ref.~\citep{BuonannoEOBv2Main} has been calibrated against higher order modes
from numerical relativity simulations. We investigate the effect of ignoring
these modes in a search by modeling the BBH signal as an EOBNRv2 signal
containing the dominant and sub-dominant multipoles: $h_{lm} =
h_{22},h_{21},h_{33},h_{44}, h_{55}$ (which we call EOBNRv2HM) and computing
the fitting factor of leading-order EOBNRv2 templates placed using the
TaylorF2 metric.

We simulate a population of BBH signals by sampling $100,000$ systems
uniformly in the $m_1,m_2 \in [3,25] \, M_\odot$ component-mass space. These
EOBNRv2HM signals are uniformly distributed in sky-location angles and
inclination and polarization angles, which appear in the detector's response
function to the gravitational-wave signal~\citep{Sathyaprakash:2009xs}. The 
template bank
is again placed with a desired minimal match of $0.97$ and for each of signal
waveforms, we calculate the $\FF$ against the entire bank of EOBNRv2 waveform
templates.  Fig.~\ref{fig:match_eob22eobhm_m1m2} (left panel) shows the value 
of the $\FF$
of the bank of EOBNRv2 waveform templates over the sampled component-mass 
space. As expected, the higest fitting factors are observed close to the equal mass
line, since when the mass ratio is close to unity, the amplitude of the
sub-leading waveform modes is several orders of magnitude smaller than the
amplitude of the dominant mode. As the mass ratio increases, the relative
amplitude of the sub-leading multipoles increases, as illustrated by Fig.~1 of
Ref.~\citep{BuonannoEOBv2Main} and the fitting factor decreases. This pattern
is brought out further in Fig.~\ref{fig:incvsM_eob22eobhm} (left panel),
where we show the $\FF$ values in the mass ratio - inclination angle 
($q-\iota$) plane. We
observe that when the orbital angular momentum is either parallel or anti-parallel
to the line of sight from the detector, the sub-leading multipoles do not 
contribute significantly to the signal. This is what we would expect from
Eq.~\ref{eq:hpcfromhlm}, as the spin-weighted harmonics are proportional to
$\mathrm{sin}(\dfrac{\iota}{2})\mathrm{cos}(\dfrac{\iota}{2})$, except when
$l=m=2$. Similar to Sec.~\ref{s2:eob22f2}, we divide the sampled component-mass
space into sub-regions bounded by $1\leq q \leq\,q_{\mathrm{threshold}}$, and compute
the minimal-fitting-factor of the EOBNRv2 template bank over those. In
Fig.~\ref{fig:match_eob22eobhm_m1m2} (right panel), the blue (solid) curve shows the 
value of $q_{\mathrm{threshold}}$ for each restricted sub-region against the 
minimal-fitting-factor of the bank over the same. For systems
with mass-ratio $q$ below $1.68\, (4)$, we find that the $\FF$ of the
EOBNRv2 waveform bank is above $0.965\, (0.947)$ over $99.75\%$ of this
restricted region.  These results demonstrate that
the effect of ignoring sub-dominant modes does not cause a significant loss in
the total possible signal-to-noise if the mass ratio is less than $1.68$. 
Similar analysis over the range of possible inclination angles shows that the EOBNRv2 
waveform bank has fitting factors above $0.965\, (0.947)$ for systems with
$2.68\,(2.02)\leq \iota \leq \pi$, and $0\leq \iota \leq 0.31\,(1.08)$ (see 
Fig.~\ref{fig:incvsM_eob22eobhm}, left panel).

Fitting factors as low as $0.92$ are observed for systems with high mass-ratios 
\textit{and} inclination angle close to $\pi/2$. As these are also binary 
configurations to which the detector is relatively less sensitive to \citep{Pekowsky:2012sr}, 
fitting-factors alone do not answer the question of where in the parameter 
space do we lose the most, in terms of detection rate. To address
this question, we compute the volume-weighted fitting-factors $\epsilon_{\mathrm{V}}$
of the EOBNRv2
template bank, over the sampled BBH parameter space. This gives us an estimate of
the expected loss in detection rate, if the source population is distributed 
uniformly in spacial volume and uniformly in intrinsic and extrinsic source parameters. 
Fig.~\ref{fig:VeffLoss_eob22eobhm} (left panel) shows $\epsilon_{\mathrm{V}}$ 
calculated in bins over the
component-mass space. In this figure, the color of each bin in the component-mass 
plane corresponds to, for a population which has all other parameters i.e. the
inclination angle and sky/polarization angles uniformly distributed over their 
possible ranges, the averaged loss in the detection rate incurred due to the use 
of a bank of leading-order $l=m=2$ EOBNRv2 templates, placed using the 2PN bank 
placement metric. We observe that
the maximum loss incurred goes up to only $\sim 10\% - 11\%$, which is within our
acceptable threshold. Looking at Fig.~\ref{fig:incvsM_eob22eobhm} (left panel), 
the maximum loss in fitting factor occurs for systems with inclination angles 
close to $\pi/2$, but (for the same mass-ratio) these
gets averaged out with systems with inclinations close to $0$ or $\pi$, which
leads to the low averaged detection-rate losses we observe in 
Fig.~\ref{fig:VeffLoss_eob22eobhm} (left panel). The right panel of 
Fig.~\ref{fig:VeffLoss_eob22eobhm} shows the same quantity, $\epsilon_{\mathrm{V}}$,
calculated over bins in the mass-ratio - inclination angle plane. As expected, 
we observe that, letting all other parameters be distributed uniformly over their
possible ranges, systems with high mass-ratios \textit{and} inclination angles 
close to $\pi/2$ will incur (averaged) losses in observation volume of 
up to $\sim 20\%$.

These results suggest that a search that includes higher order modes
could achieve a non-trivial increase in sensitivity over leading-order mode
templates, only in detecting systems with high $q$ \textit{and} $1.08\leq\iota\leq 2.02$.  
However, an algorithm that includes sub-dominant modes could have an 
increased false-alarm rate (background) over a search that includes 
only the leading-order mode, and hence the overall gain in search efficiency
might not be significant. 

\section{Conclusions}
\label{s:conclusions}

We used the TaylorF2 second-order post-Newtonian hexagonal placement algorithm
of Refs.~\citep{SathyaMetric2PN,BabaketalBankPlacement,Cokelaer:2007kx} to construct a
template bank of EOBNRv2 waveforms with $\MM$ of $0.97$. We calculated the
fitting factor ($\FF$) of this bank against $\sim 90,000$ simulated EOBNRv2
signals with component masses uniformly distributed between $3 M_{\odot} \le
m_1,m_2 \le 25 M_{\odot}$. We find that the $\FF$ of the template bank is
greater than $0.97$ for $98.5\%$ of the simulated EOBNRv2 signals, assuming
the zero-detuning high power noise spectrum for aLIGO sensitivity
\citep{aLIGONoiseCurve}. We conclude that the existing placement algorithm is
effectual for use in aLIGO BBH searches, assuming that EOBNRv2 is an accurate
model of BBH signals in this mass region.  We then demonstrated that use of the
computationally cheaper third-and-a-half order TaylorF2 waveform results in a
loss in search efficiency due to inaccuracies of the post-Newtonian
approximation, and neglect of merger-ringdown for BBHs with a total mass $M >
11.4\,M_{\odot}$. However, below this limit the TaylorF2 model is an acceptable
signal for BBH serches. This was done using a bank with a $\MM$ of 0.97. By 
increasing the density of the bank to $0.99\MM$, the limit on total-mass can be 
relaxed to $16.3\,M_{\odot}$, with an increase in computational cost due to the 
number of templates increasing by a factor of $\sim 2.7$. Finally, we investigated 
the loss in the SNR incurred by the using template banks constructed using 
only the leading order mode of EOBNRv2 waveforms.  We found that a leading-order 
$l=m=2$ EOBNRv2 template bank constructed with a $\MM$ of $0.97$ is effectual 
to search for BBHs for which $1\leq \left(m_1/m_2\right)\leq 1.68$ or $\iota \geq 2.68$ 
or $\iota\leq 0.31$ radians, and there is no significant loss in
potential signal-to-noise ratio for systems with $q$ as high as $4$ or
$2.02\leq\iota\leq 2.68$ or $0.31\leq\iota\leq 1.08$. We also observed that the
maximum loss in detection-rate, for a binary with given mass parameters, averaging
over other parameters - which are taken to be uniformly distributed over their 
possible ranges, goes only to a maximum of $\sim 10\% - 11\%$. For any given pair of 
binary masses, the loss is highest when the binary is inclined at $\simeq \pi/2$, 
and can go up to $\sim 20\%$, and is lower when its angular momentum is close to 
being parallel or anti-parallel to the line of sight from the detector. These effects 
average out, and hence for a population which is expected to have a uniform 
distribution of inclination angles (and uniform distribution in spacial volume), 
the average loss in detection rate was estimated to be not higher than $\sim 11\%$.
Thus, using EOBNRv2HM templates is unlikely to give a significant increase
in the range to which such a population of sources can be detected. For BBHs 
with $\left(m_1/m_2\right)\geq 4$ \textit{and} $1.08\leq\iota\leq 2.02$, detection searches 
could possibly gain sensitivity by the use of EOBNRv2HM waveforms, if they can be implemented 
without increasing the false alarm rate. 

Our results suggest that a significant portion of the non-spinning
stellar-mass BBH parameter space can be searched for using LIGO's existing
search algorithms. For systems with total mass below $\sim 11.4M_{\odot}$ template
banks of TaylorF2 can be used without significant loss in event rate. For
higher mass systems, neglecting high-order modes in an EOBNRv2 search does not
cause a substantial reduction in the maximum possible reach of BBH searches.
Finally, we note that our study does not consider BBH systems with BH
masses higher than $M = 25\,M_\odot$, or the effect of black hole component
spins. Future work will extend this study for systems with spinning and/or
precessing black holes and consider the effect of non-Gaussian transients in
real detector noise.

\acknowledgments
We are grateful to Stefan Ballmer, Jolien Creighton, Ian Harry, Eliu Huerta,
Peter Saulson, and Matt West for helpful comments. We thank Tom Dent for
carefully reading this manuscript and Yi Pan for fixing a problem with the
EOBNRv2 model discovered during this work.  We are grateful to Evan Ochsner
and Craig Robinson for writing the LAL implementation of EOBNRv2 against which
our code was validated. We are also grateful to the anonymous peer-review
referee for their constructive comments. This work was supported NSF grants 
PHY-0847611 and PHY-0854812, and a Research Corporation for Science Advancement 
Cottrell Scholar award.  Computations were carried out on the SUGAR cluster 
which is supported by NSF grants PHY-1040231, PHY-1104371, and PHY-0600953, 
and by Syracuse University ITS. Part of this work was carried out at the Kavli
Institute for Theoretical Physics at Santa Barbara University, supported in
part by NSF grant PHY-0551164.


\begin{thebibliography}{69}%
\makeatletter
\providecommand \@ifxundefined [1]{%
 \@ifx{#1\undefined}
}%
\providecommand \@ifnum [1]{%
 \ifnum #1\expandafter \@firstoftwo
 \else \expandafter \@secondoftwo
 \fi
}%
\providecommand \@ifx [1]{%
 \ifx #1\expandafter \@firstoftwo
 \else \expandafter \@secondoftwo
 \fi
}%
\providecommand \natexlab [1]{#1}%
\providecommand \enquote  [1]{``#1''}%
\providecommand \bibnamefont  [1]{#1}%
\providecommand \bibfnamefont [1]{#1}%
\providecommand \citenamefont [1]{#1}%
\providecommand \href@noop [0]{\@secondoftwo}%
\providecommand \href [0]{\begingroup \@sanitize@url \@href}%
\providecommand \@href[1]{\@@startlink{#1}\@@href}%
\providecommand \@@href[1]{\endgroup#1\@@endlink}%
\providecommand \@sanitize@url [0]{\catcode `\\12\catcode `\$12\catcode
  `\&12\catcode `\#12\catcode `\^12\catcode `\_12\catcode `\%12\relax}%
\providecommand \@@startlink[1]{}%
\providecommand \@@endlink[0]{}%
\providecommand \url  [0]{\begingroup\@sanitize@url \@url }%
\providecommand \@url [1]{\endgroup\@href {#1}{\urlprefix }}%
\providecommand \urlprefix  [0]{URL }%
\providecommand \Eprint [0]{\href }%
\providecommand \doibase [0]{http://dx.doi.org/}%
\providecommand \selectlanguage [0]{\@gobble}%
\providecommand \bibinfo  [0]{\@secondoftwo}%
\providecommand \bibfield  [0]{\@secondoftwo}%
\providecommand \translation [1]{[#1]}%
\providecommand \BibitemOpen [0]{}%
\providecommand \bibitemStop [0]{}%
\providecommand \bibitemNoStop [0]{.\EOS\space}%
\providecommand \EOS [0]{\spacefactor3000\relax}%
\providecommand \BibitemShut  [1]{\csname bibitem#1\endcsname}%
\let\auto@bib@innerbib\@empty
\bibitem [{\citenamefont {Harry}\ \emph {et~al.}(2010)\citenamefont {Harry}
  \emph {et~al.}}]{Harry:2010zz}%
  \BibitemOpen
  \bibfield  {author} {\bibinfo {author} {\bibfnamefont {G.~M.}\ \bibnamefont
  {Harry}} \emph {et~al.} (\bibinfo {collaboration} {LIGO Scientific
  Collaboration}),\ }\href {\doibase 10.1088/0264-9381/27/8/084006} {\bibfield
  {journal} {\bibinfo  {journal} {Class.Quant.Grav.}\ }\textbf {\bibinfo
  {volume} {27}},\ \bibinfo {pages} {084006} (\bibinfo {year}
  {2010})}\BibitemShut {NoStop}%
\bibitem [{\citenamefont {Acernese}\ \emph {et~al.}(2009)\citenamefont
  {Acernese} \emph {et~al.}}]{aVIRGO}%
  \BibitemOpen
  \bibfield  {author} {\bibinfo {author} {\bibfnamefont {F.}~\bibnamefont
  {Acernese}} \emph {et~al.} (\bibinfo {collaboration} {{The Virgo
  Collaboration}}),\ }\href@noop {} {\enquote {\bibinfo {title} {{Advanced
  Virgo Baseline Design}},}\ } (\bibinfo {year} {2009}),\ \bibinfo {note}
  {{[Virgo Techincal Document VIR-0027A-09]}}\BibitemShut {NoStop}%
\bibitem [{\citenamefont {Somiya}(2012)}]{Somiya:2011np}%
  \BibitemOpen
  \bibfield  {author} {\bibinfo {author} {\bibfnamefont {K.}~\bibnamefont
  {Somiya}} (\bibinfo {collaboration} {KAGRA Collaboration}),\ }\href {\doibase
  10.1088/0264-9381/29/12/124007} {\bibfield  {journal} {\bibinfo  {journal}
  {Class.Quant.Grav.}\ }\textbf {\bibinfo {volume} {29}},\ \bibinfo {pages}
  {124007} (\bibinfo {year} {2012})},\ \Eprint {http://arxiv.org/abs/1111.7185}
  {arXiv:1111.7185 [gr-qc]} \BibitemShut {NoStop}%
\bibitem [{\citenamefont {Waldman}(2011)}]{aLIGOsensitivity}%
  \BibitemOpen
  \bibfield  {author} {\bibinfo {author} {\bibfnamefont {S.}~\bibnamefont
  {Waldman}} (\bibinfo {collaboration} {the LIGO Scientific Collaboration}),\
  }\href@noop {} {\  (\bibinfo {year} {2011})},\ \Eprint
  {http://arxiv.org/abs/1103.2728} {arXiv:1103.2728 [gr-qc]} \BibitemShut
  {NoStop}%
\bibitem [{\citenamefont {Hawking}\ and\ \citenamefont
  {Israel}(1987)}]{300yrsofGravitation}%
  \BibitemOpen
  \bibfield  {author} {\bibinfo {author} {\bibfnamefont {S.~W.}\ \bibnamefont
  {Hawking}}\ and\ \bibinfo {author} {\bibfnamefont {W.}~\bibnamefont
  {Israel}},\ }\href@noop {} {\emph {\bibinfo {title} {Three Hundred Years of
  Gravitation}}}\ (\bibinfo  {publisher} {Cambridge University Press},\
  \bibinfo {year} {1987})\BibitemShut {NoStop}%
\bibitem [{\citenamefont {Abadie}\ \emph
  {et~al.}(2010{\natexlab{a}})\citenamefont {Abadie} \emph
  {et~al.}}]{LSCCBCRates2010}%
  \BibitemOpen
  \bibfield  {author} {\bibinfo {author} {\bibfnamefont {J.}~\bibnamefont
  {Abadie}} \emph {et~al.} (\bibinfo {collaboration} {LIGO Scientific
  Collaboration, Virgo Collaboration}),\ }\href {\doibase
  10.1088/0264-9381/27/17/173001} {\bibfield  {journal} {\bibinfo  {journal}
  {Class.Quant.Grav.}\ }\textbf {\bibinfo {volume} {27}},\ \bibinfo {pages}
  {173001} (\bibinfo {year} {2010}{\natexlab{a}})},\ \Eprint
  {http://arxiv.org/abs/1003.2480} {arXiv:1003.2480 [astro-ph.HE]} \BibitemShut
  {NoStop}%
\bibitem [{\citenamefont {Buonanno}\ and\ \citenamefont
  {Damour}(1999{\natexlab{a}})}]{EOBOriginalBuonannoDamour}%
  \BibitemOpen
  \bibfield  {author} {\bibinfo {author} {\bibfnamefont {A.}~\bibnamefont
  {Buonanno}}\ and\ \bibinfo {author} {\bibfnamefont {T.}~\bibnamefont
  {Damour}},\ }\href {\doibase 10.1103/PhysRevD.59.084006} {\bibfield
  {journal} {\bibinfo  {journal} {Phys.Rev.}\ }\textbf {\bibinfo {volume}
  {D59}},\ \bibinfo {pages} {084006} (\bibinfo {year} {1999}{\natexlab{a}})},\
  \Eprint {http://arxiv.org/abs/gr-qc/9811091} {arXiv:gr-qc/9811091 [gr-qc]}
  \BibitemShut {NoStop}%
\bibitem [{\citenamefont {Damour}\ \emph
  {et~al.}(2008{\natexlab{a}})\citenamefont {Damour}, \citenamefont {Nagar},
  \citenamefont {Hannam}, \citenamefont {Husa},\ and\ \citenamefont
  {Brugmann}}]{EOBNR01}%
  \BibitemOpen
  \bibfield  {author} {\bibinfo {author} {\bibfnamefont {T.}~\bibnamefont
  {Damour}}, \bibinfo {author} {\bibfnamefont {A.}~\bibnamefont {Nagar}},
  \bibinfo {author} {\bibfnamefont {M.}~\bibnamefont {Hannam}}, \bibinfo
  {author} {\bibfnamefont {S.}~\bibnamefont {Husa}}, \ and\ \bibinfo {author}
  {\bibfnamefont {B.}~\bibnamefont {Brugmann}},\ }\href {\doibase
  10.1103/PhysRevD.78.044039} {\bibfield  {journal} {\bibinfo  {journal}
  {Phys.Rev.}\ }\textbf {\bibinfo {volume} {D78}},\ \bibinfo {pages} {044039}
  (\bibinfo {year} {2008}{\natexlab{a}})},\ \Eprint
  {http://arxiv.org/abs/0803.3162} {arXiv:0803.3162 [gr-qc]} \BibitemShut
  {NoStop}%
\bibitem [{\citenamefont {Buonanno}\ \emph
  {et~al.}(2009{\natexlab{a}})\citenamefont {Buonanno}, \citenamefont {Pan},
  \citenamefont {Pfeiffer}, \citenamefont {Scheel}, \citenamefont {Buchman}
  \emph {et~al.}}]{EOBNRdevel01}%
  \BibitemOpen
  \bibfield  {author} {\bibinfo {author} {\bibfnamefont {A.}~\bibnamefont
  {Buonanno}}, \bibinfo {author} {\bibfnamefont {Y.}~\bibnamefont {Pan}},
  \bibinfo {author} {\bibfnamefont {H.~P.}\ \bibnamefont {Pfeiffer}}, \bibinfo
  {author} {\bibfnamefont {M.~A.}\ \bibnamefont {Scheel}}, \bibinfo {author}
  {\bibfnamefont {L.~T.}\ \bibnamefont {Buchman}},  \emph {et~al.},\ }\href
  {\doibase 10.1103/PhysRevD.79.124028} {\bibfield  {journal} {\bibinfo
  {journal} {Phys.Rev.}\ }\textbf {\bibinfo {volume} {D79}},\ \bibinfo {pages}
  {124028} (\bibinfo {year} {2009}{\natexlab{a}})},\ \Eprint
  {http://arxiv.org/abs/0902.0790} {arXiv:0902.0790 [gr-qc]} \BibitemShut
  {NoStop}%
\bibitem [{\citenamefont {Damour}\ and\ \citenamefont
  {Nagar}(2009)}]{EOBNRdevel02}%
  \BibitemOpen
  \bibfield  {author} {\bibinfo {author} {\bibfnamefont {T.}~\bibnamefont
  {Damour}}\ and\ \bibinfo {author} {\bibfnamefont {A.}~\bibnamefont {Nagar}},\
  }\href {\doibase 10.1103/PhysRevD.79.081503} {\bibfield  {journal} {\bibinfo
  {journal} {Phys.Rev.}\ }\textbf {\bibinfo {volume} {D79}},\ \bibinfo {pages}
  {081503} (\bibinfo {year} {2009})},\ \Eprint {http://arxiv.org/abs/0902.0136}
  {arXiv:0902.0136 [gr-qc]} \BibitemShut {NoStop}%
\bibitem [{\citenamefont {Damour}\ and\ \citenamefont
  {Nagar}(2008)}]{EOBNRdevel03}%
  \BibitemOpen
  \bibfield  {author} {\bibinfo {author} {\bibfnamefont {T.}~\bibnamefont
  {Damour}}\ and\ \bibinfo {author} {\bibfnamefont {A.}~\bibnamefont {Nagar}},\
  }\href {\doibase 10.1103/PhysRevD.77.024043} {\bibfield  {journal} {\bibinfo
  {journal} {Phys.Rev.}\ }\textbf {\bibinfo {volume} {D77}},\ \bibinfo {pages}
  {024043} (\bibinfo {year} {2008})},\ \Eprint {http://arxiv.org/abs/0711.2628}
  {arXiv:0711.2628 [gr-qc]} \BibitemShut {NoStop}%
\bibitem [{\citenamefont {Damour}\ \emph
  {et~al.}(2008{\natexlab{b}})\citenamefont {Damour}, \citenamefont {Nagar},
  \citenamefont {Dorband}, \citenamefont {Pollney},\ and\ \citenamefont
  {Rezzolla}}]{EOBNRdevel04}%
  \BibitemOpen
  \bibfield  {author} {\bibinfo {author} {\bibfnamefont {T.}~\bibnamefont
  {Damour}}, \bibinfo {author} {\bibfnamefont {A.}~\bibnamefont {Nagar}},
  \bibinfo {author} {\bibfnamefont {E.~N.}\ \bibnamefont {Dorband}}, \bibinfo
  {author} {\bibfnamefont {D.}~\bibnamefont {Pollney}}, \ and\ \bibinfo
  {author} {\bibfnamefont {L.}~\bibnamefont {Rezzolla}},\ }\href {\doibase
  10.1103/PhysRevD.77.084017} {\bibfield  {journal} {\bibinfo  {journal}
  {Phys.Rev.}\ }\textbf {\bibinfo {volume} {D77}},\ \bibinfo {pages} {084017}
  (\bibinfo {year} {2008}{\natexlab{b}})},\ \Eprint
  {http://arxiv.org/abs/0712.3003} {arXiv:0712.3003 [gr-qc]} \BibitemShut
  {NoStop}%
\bibitem [{\citenamefont {Buonanno}\ and\ \citenamefont
  {Damour}(2000)}]{EOBdevel01}%
  \BibitemOpen
  \bibfield  {author} {\bibinfo {author} {\bibfnamefont {A.}~\bibnamefont
  {Buonanno}}\ and\ \bibinfo {author} {\bibfnamefont {T.}~\bibnamefont
  {Damour}},\ }\href {\doibase 10.1103/PhysRevD.62.064015} {\bibfield
  {journal} {\bibinfo  {journal} {Phys.Rev.}\ }\textbf {\bibinfo {volume}
  {D62}},\ \bibinfo {pages} {064015} (\bibinfo {year} {2000})},\ \bibinfo
  {note} {52 pages, 21 figures, ReVTex, epsfig; few misprints corrected},\
  \Eprint {http://arxiv.org/abs/gr-qc/0001013} {arXiv:gr-qc/0001013 [gr-qc]}
  \BibitemShut {NoStop}%
\bibitem [{\citenamefont {Damour}\ \emph {et~al.}(2003)\citenamefont {Damour},
  \citenamefont {Iyer}, \citenamefont {Jaranowski},\ and\ \citenamefont
  {Sathyaprakash}}]{EOBdevel02}%
  \BibitemOpen
  \bibfield  {author} {\bibinfo {author} {\bibfnamefont {T.}~\bibnamefont
  {Damour}}, \bibinfo {author} {\bibfnamefont {B.~R.}\ \bibnamefont {Iyer}},
  \bibinfo {author} {\bibfnamefont {P.}~\bibnamefont {Jaranowski}}, \ and\
  \bibinfo {author} {\bibfnamefont {B.~S.}~\bibnamefont {Sathyaprakash}},\ }\href
  {\doibase 10.1103/PhysRevD.67.064028} {\bibfield  {journal} {\bibinfo
  {journal} {Phys.Rev.}\ }\textbf {\bibinfo {volume} {D67}},\ \bibinfo {pages}
  {064028} (\bibinfo {year} {2003})},\ \Eprint
  {http://arxiv.org/abs/gr-qc/0211041} {arXiv:gr-qc/0211041 [gr-qc]}
  \BibitemShut {NoStop}%
\bibitem [{\citenamefont {Pan}\ \emph {et~al.}(2011)\citenamefont {Pan},
  \citenamefont {Buonanno}, \citenamefont {Boyle}, \citenamefont {Buchman},
  \citenamefont {Kidder} \emph {et~al.}}]{BuonannoEOBv2Main}%
  \BibitemOpen
  \bibfield  {author} {\bibinfo {author} {\bibfnamefont {Y.}~\bibnamefont
  {Pan}}, \bibinfo {author} {\bibfnamefont {A.}~\bibnamefont {Buonanno}},
  \bibinfo {author} {\bibfnamefont {M.}~\bibnamefont {Boyle}}, \bibinfo
  {author} {\bibfnamefont {L.~T.}\ \bibnamefont {Buchman}}, \bibinfo {author}
  {\bibfnamefont {L.~E.}\ \bibnamefont {Kidder}},  \emph {et~al.},\ }\href
  {\doibase 10.1103/PhysRevD.84.124052} {\bibfield  {journal} {\bibinfo
  {journal} {Phys.Rev.}\ }\textbf {\bibinfo {volume} {D84}},\ \bibinfo {pages}
  {124052} (\bibinfo {year} {2011})},\ \bibinfo {note} {26 pages, 25 figures},\
  \Eprint {http://arxiv.org/abs/1106.1021} {arXiv:1106.1021 [gr-qc]}
  \BibitemShut {NoStop}%
\bibitem [{\citenamefont {Berti}\ \emph {et~al.}(2006)\citenamefont {Berti},
  \citenamefont {Cardoso},\ and\ \citenamefont {Will}}]{BHRDQNMs}%
  \BibitemOpen
  \bibfield  {author} {\bibinfo {author} {\bibfnamefont {E.}~\bibnamefont
  {Berti}}, \bibinfo {author} {\bibfnamefont {V.}~\bibnamefont {Cardoso}}, \
  and\ \bibinfo {author} {\bibfnamefont {C.~M.}\ \bibnamefont {Will}},\ }\href
  {\doibase 10.1103/PhysRevD.73.064030} {\bibfield  {journal} {\bibinfo
  {journal} {Phys.Rev.}\ }\textbf {\bibinfo {volume} {D73}},\ \bibinfo {pages}
  {064030} (\bibinfo {year} {2006})},\ \Eprint
  {http://arxiv.org/abs/gr-qc/0512160} {arXiv:gr-qc/0512160 [gr-qc]}
  \BibitemShut {NoStop}%
\bibitem [{\citenamefont {Mino}\ \emph {et~al.}(1997)\citenamefont {Mino},
  \citenamefont {Sasaki}, \citenamefont {Shibata}, \citenamefont {Tagoshi},\
  and\ \citenamefont {Tanaka}}]{BHPTMinoSasaki}%
  \BibitemOpen
  \bibfield  {author} {\bibinfo {author} {\bibfnamefont {Y.}~\bibnamefont
  {Mino}}, \bibinfo {author} {\bibfnamefont {M.}~\bibnamefont {Sasaki}},
  \bibinfo {author} {\bibfnamefont {M.}~\bibnamefont {Shibata}}, \bibinfo
  {author} {\bibfnamefont {H.}~\bibnamefont {Tagoshi}}, \ and\ \bibinfo
  {author} {\bibfnamefont {T.}~\bibnamefont {Tanaka}},\ }\href {\doibase
  10.1143/PTPS.128.1} {\bibfield  {journal} {\bibinfo  {journal}
  {Prog.Theor.Phys.Suppl.}\ }\textbf {\bibinfo {volume} {128}},\ \bibinfo
  {pages} {1} (\bibinfo {year} {1997})},\ \Eprint
  {http://arxiv.org/abs/gr-qc/9712057} {arXiv:gr-qc/9712057 [gr-qc]}
  \BibitemShut {NoStop}%
\bibitem [{\citenamefont {Abadie}\ \emph {et~al.}(2012)\citenamefont {Abadie}
  \emph {et~al.}}]{Colaboration:2011nz}%
  \BibitemOpen
  \bibfield  {author} {\bibinfo {author} {\bibfnamefont {J.}~\bibnamefont
  {Abadie}} \emph {et~al.} (\bibinfo {collaboration} {LIGO Scientific
  Collaboration, Virgo Collaboration}),\ }\href {\doibase
  10.1103/PhysRevD.85.082002} {\bibfield  {journal} {\bibinfo  {journal}
  {Phys.Rev.}\ }\textbf {\bibinfo {volume} {D85}},\ \bibinfo {pages} {082002}
  (\bibinfo {year} {2012})},\ \Eprint {http://arxiv.org/abs/1111.7314}
  {arXiv:1111.7314 [gr-qc]} \BibitemShut {NoStop}%
\bibitem [{\citenamefont {Abadie}\ \emph
  {et~al.}(2010{\natexlab{b}})\citenamefont {Abadie} \emph
  {et~al.}}]{Abadie:2010yb}%
  \BibitemOpen
  \bibfield  {author} {\bibinfo {author} {\bibfnamefont {J.}~\bibnamefont
  {Abadie}} \emph {et~al.} (\bibinfo {collaboration} {LIGO Scientific
  Collaboration, Virgo Collaboration}),\ }\href {\doibase
  10.1103/PhysRevD.85.089903, 10.1103/PhysRevD.82.102001} {\bibfield  {journal}
  {\bibinfo  {journal} {Phys.Rev.}\ }\textbf {\bibinfo {volume} {D82}},\
  \bibinfo {pages} {102001} (\bibinfo {year} {2010}{\natexlab{b}})},\ \Eprint
  {http://arxiv.org/abs/1005.4655} {arXiv:1005.4655 [gr-qc]} \BibitemShut
  {NoStop}%
\bibitem [{\citenamefont {Abbott}\ \emph
  {et~al.}(2009{\natexlab{a}})\citenamefont {Abbott} \emph
  {et~al.}}]{Abbott:2009qj}%
  \BibitemOpen
  \bibfield  {author} {\bibinfo {author} {\bibfnamefont {B.}~\bibnamefont
  {Abbott}} \emph {et~al.} (\bibinfo {collaboration} {LIGO Scientific
  Collaboration}),\ }\href {\doibase 10.1103/PhysRevD.80.047101} {\bibfield
  {journal} {\bibinfo  {journal} {Phys.Rev.}\ }\textbf {\bibinfo {volume}
  {D80}},\ \bibinfo {pages} {047101} (\bibinfo {year} {2009}{\natexlab{a}})},\
  \Eprint {http://arxiv.org/abs/0905.3710} {arXiv:0905.3710 [gr-qc]}
  \BibitemShut {NoStop}%
\bibitem [{\citenamefont {Abbott}\ \emph
  {et~al.}(2009{\natexlab{b}})\citenamefont {Abbott} \emph
  {et~al.}}]{Abbott:2009tt}%
  \BibitemOpen
  \bibfield  {author} {\bibinfo {author} {\bibfnamefont {B.}~\bibnamefont
  {Abbott}} \emph {et~al.} (\bibinfo {collaboration} {LIGO Scientific
  Collaboration}),\ }\href {\doibase 10.1103/PhysRevD.79.122001} {\bibfield
  {journal} {\bibinfo  {journal} {Phys.Rev.}\ }\textbf {\bibinfo {volume}
  {D79}},\ \bibinfo {pages} {122001} (\bibinfo {year} {2009}{\natexlab{b}})},\
  \Eprint {http://arxiv.org/abs/0901.0302} {arXiv:0901.0302 [gr-qc]}
  \BibitemShut {NoStop}%
\bibitem [{\citenamefont {Messaritaki}(2005)}]{Messaritaki:2005wv}%
  \BibitemOpen
  \bibfield  {author} {\bibinfo {author} {\bibfnamefont {E.}~\bibnamefont
  {Messaritaki}} (\bibinfo {collaboration} {LIGO Scientific Collaboration}),\
  }\href {\doibase 10.1088/0264-9381/22/18/S26} {\bibfield  {journal} {\bibinfo
   {journal} {Class.Quant.Grav.}\ }\textbf {\bibinfo {volume} {22}},\ \bibinfo
  {pages} {S1119} (\bibinfo {year} {2005})},\ \Eprint
  {http://arxiv.org/abs/gr-qc/0504065} {arXiv:gr-qc/0504065 [gr-qc]}
  \BibitemShut {NoStop}%
\bibitem [{\citenamefont {Wainstein}\ and\ \citenamefont
  {Zubakov}(1962)}]{Wainstein:1962}%
  \BibitemOpen
  \bibfield  {author} {\bibinfo {author} {\bibfnamefont {L.~A.}\ \bibnamefont
  {Wainstein}}\ and\ \bibinfo {author} {\bibfnamefont {V.~D.}\ \bibnamefont
  {Zubakov}},\ }\href@noop {} {\emph {\bibinfo {title} {Extraction of signals
  from noise}}}\ (\bibinfo  {publisher} {Prentice-Hall},\ \bibinfo {address}
  {Englewood Cliffs, NJ},\ \bibinfo {year} {1962})\BibitemShut {NoStop}%
\bibitem [{\citenamefont {Allen}\ \emph {et~al.}(2005)\citenamefont {Allen}
  \emph {et~al.}}]{Allen:2005fk}%
  \BibitemOpen
  \bibfield  {author} {\bibinfo {author} {\bibfnamefont {B.}~\bibnamefont
  {Allen}} \emph {et~al.},\ }\href@noop {} {\  (\bibinfo {year} {2005})},\
  \Eprint {http://arxiv.org/abs/gr-qc/0509116} {arXiv:gr-qc/0509116 [gr-qc]}
  \BibitemShut {NoStop}%
\bibitem [{\citenamefont {Ozel}\ \emph {et~al.}(2010)\citenamefont {Ozel},
  \citenamefont {Psaltis}, \citenamefont {Narayan},\ and\ \citenamefont
  {McClintock}}]{Ozel:2010su}%
  \BibitemOpen
  \bibfield  {author} {\bibinfo {author} {\bibfnamefont {F.}~\bibnamefont
  {Ozel}}, \bibinfo {author} {\bibfnamefont {D.}~\bibnamefont {Psaltis}},
  \bibinfo {author} {\bibfnamefont {R.}~\bibnamefont {Narayan}}, \ and\
  \bibinfo {author} {\bibfnamefont {J.~E.}\ \bibnamefont {McClintock}},\ }\href
  {\doibase 10.1088/0004-637X/725/2/1918} {\bibfield  {journal} {\bibinfo
  {journal} {Astrophys.J.}\ }\textbf {\bibinfo {volume} {725}},\ \bibinfo
  {pages} {1918} (\bibinfo {year} {2010})},\ \Eprint
  {http://arxiv.org/abs/1006.2834} {arXiv:1006.2834 [astro-ph.GA]} \BibitemShut
  {NoStop}%
\bibitem [{\citenamefont {Sathyaprakash}\ and\ \citenamefont
  {Dhurandhar}(1991)}]{Sathyaprakash:1991mt}%
  \BibitemOpen
  \bibfield  {author} {\bibinfo {author} {\bibfnamefont {B.~S.}~\bibnamefont
  {Sathyaprakash}}\ and\ \bibinfo {author} {\bibfnamefont {S.~V.}~\bibnamefont
  {Dhurandhar}},\ }\href {\doibase 10.1103/PhysRevD.44.3819} {\bibfield
  {journal} {\bibinfo  {journal} {Phys.Rev.}\ }\textbf {\bibinfo {volume}
  {D44}},\ \bibinfo {pages} {3819} (\bibinfo {year} {1991})}\BibitemShut
  {NoStop}%
\bibitem [{\citenamefont {Balasubramanian}\ \emph {et~al.}(1996)\citenamefont
  {Balasubramanian}, \citenamefont {Sathyaprakash},\ and\ \citenamefont
  {Dhurandhar}}]{Balasubramanian:1995bm}%
  \BibitemOpen
  \bibfield  {author} {\bibinfo {author} {\bibfnamefont {R.}~\bibnamefont
  {Balasubramanian}}, \bibinfo {author} {\bibfnamefont {B.~S.}~\bibnamefont
  {Sathyaprakash}}, \ and\ \bibinfo {author} {\bibfnamefont {S.~V.}~\bibnamefont
  {Dhurandhar}},\ }\href {\doibase 10.1103/PhysRevD.54.1860,
  10.1103/PhysRevD.53.3033} {\bibfield  {journal} {\bibinfo  {journal}
  {Phys.Rev.}\ }\textbf {\bibinfo {volume} {D53}},\ \bibinfo {pages} {3033}
  (\bibinfo {year} {1996})},\ \Eprint {http://arxiv.org/abs/gr-qc/9508011}
  {arXiv:gr-qc/9508011 [gr-qc]} \BibitemShut {NoStop}%
\bibitem [{\citenamefont {Apostolatos}(1995)}]{FittingFactorApostolatos}%
  \BibitemOpen
  \bibfield  {author} {\bibinfo {author} {\bibfnamefont {T.~A.}~\bibnamefont
  {Apostolatos}},\ }\href {\doibase 10.1103/PhysRevD.52.605} {\bibfield
  {journal} {\bibinfo  {journal} {Phys.Rev.}\ }\textbf {\bibinfo {volume}
  {D52}},\ \bibinfo {pages} {605} (\bibinfo {year} {1995})}\BibitemShut
  {NoStop}%
\bibitem [{\citenamefont {Cannon}\ \emph {et~al.}(2010)\citenamefont {Cannon}
  \emph {et~al.}}]{Cannon:2010qh}%
  \BibitemOpen
  \bibfield  {author} {\bibinfo {author} {\bibfnamefont {K.}~\bibnamefont
  {Cannon}} \emph {et~al.},\ }\href@noop {} {\bibfield  {journal} {\bibinfo
  {journal} {Phys.Rev.}\ }\textbf {\bibinfo {volume} {D82}},\ \bibinfo {pages}
  {044025} (\bibinfo {year} {2010})}\BibitemShut {NoStop}%
\bibitem [{\citenamefont {Owen}(1996)}]{OwenTemplateSpacing}%
  \BibitemOpen
  \bibfield  {author} {\bibinfo {author} {\bibfnamefont {B.~J.}\ \bibnamefont
  {Owen}},\ }\href {\doibase 10.1103/PhysRevD.53.6749} {\bibfield  {journal}
  {\bibinfo  {journal} {Phys.Rev.}\ }\textbf {\bibinfo {volume} {D53}},\
  \bibinfo {pages} {6749} (\bibinfo {year} {1996})},\ \Eprint
  {http://arxiv.org/abs/gr-qc/9511032} {arXiv:gr-qc/9511032 [gr-qc]}
  \BibitemShut {NoStop}%
\bibitem [{\citenamefont {Sathyaprakash}(1994)}]{SathyaBankPlacementTauN}%
  \BibitemOpen
  \bibfield  {author} {\bibinfo {author} {\bibfnamefont {B.~S.}~\bibnamefont
  {Sathyaprakash}},\ }\href {\doibase 10.1103/PhysRevD.50.R7111} {\bibfield
  {journal} {\bibinfo  {journal} {Phys.Rev.}\ }\textbf {\bibinfo {volume}
  {D50}},\ \bibinfo {pages} {7111} (\bibinfo {year} {1994})},\ \Eprint
  {http://arxiv.org/abs/gr-qc/9411043} {arXiv:gr-qc/9411043 [gr-qc]}
  \BibitemShut {NoStop}%
\bibitem [{\citenamefont {Babak}\ \emph {et~al.}(2006)\citenamefont {Babak},
  \citenamefont {Balasubramanian}, \citenamefont {Churches}, \citenamefont
  {Cokelaer},\ and\ \citenamefont {Sathyaprakash}}]{BabaketalBankPlacement}%
  \BibitemOpen
  \bibfield  {author} {\bibinfo {author} {\bibfnamefont {S.}~\bibnamefont
  {Babak}}, \bibinfo {author} {\bibfnamefont {R.}~\bibnamefont
  {Balasubramanian}}, \bibinfo {author} {\bibfnamefont {D.}~\bibnamefont
  {Churches}}, \bibinfo {author} {\bibfnamefont {T.}~\bibnamefont {Cokelaer}},
  \ and\ \bibinfo {author} {\bibfnamefont {B.}~\bibnamefont {Sathyaprakash}},\
  }\href {\doibase 10.1088/0264-9381/23/18/002} {\bibfield  {journal} {\bibinfo
   {journal} {Class.Quant.Grav.}\ }\textbf {\bibinfo {volume} {23}},\ \bibinfo
  {pages} {5477} (\bibinfo {year} {2006})},\ \Eprint
  {http://arxiv.org/abs/gr-qc/0604037} {arXiv:gr-qc/0604037 [gr-qc]}
  \BibitemShut {NoStop}%
\bibitem [{\citenamefont {Owen}\ and\ \citenamefont
  {Sathyaprakash}(1999)}]{SathyaMetric2PN}%
  \BibitemOpen
  \bibfield  {author} {\bibinfo {author} {\bibfnamefont {B.~J.}\ \bibnamefont
  {Owen}}\ and\ \bibinfo {author} {\bibfnamefont {B.~S.}~\bibnamefont
  {Sathyaprakash}},\ }\href {\doibase 10.1103/PhysRevD.60.022002} {\bibfield
  {journal} {\bibinfo  {journal} {Phys.Rev.}\ }\textbf {\bibinfo {volume}
  {D60}},\ \bibinfo {pages} {022002} (\bibinfo {year} {1999})},\ \Eprint
  {http://arxiv.org/abs/gr-qc/9808076} {arXiv:gr-qc/9808076 [gr-qc]}
  \BibitemShut {NoStop}%
\bibitem [{\citenamefont {Cokelaer}(2007)}]{Cokelaer:2007kx}%
  \BibitemOpen
  \bibfield  {author} {\bibinfo {author} {\bibfnamefont {T.}~\bibnamefont
  {Cokelaer}},\ }\href {\doibase 10.1103/PhysRevD.76.102004} {\bibfield
  {journal} {\bibinfo  {journal} {Phys.Rev.}\ }\textbf {\bibinfo {volume}
  {D76}},\ \bibinfo {pages} {102004} (\bibinfo {year} {2007})},\ \Eprint
  {http://arxiv.org/abs/0706.4437} {arXiv:0706.4437 [gr-qc]} \BibitemShut
  {NoStop}%
\bibitem [{\citenamefont {{LIGO (David Shoemaker)}}(2009)}]{aLIGONoiseCurve}%
  \BibitemOpen
  \bibfield  {author} {\bibinfo {author} {\bibnamefont {{LIGO (David
  Shoemaker)}}},\ }\href@noop {} {\emph {\bibinfo {title} {{Advanced LIGO
  anticipated sensitivity curves}}}},\ \bibinfo {type} {Tech. Rep.}\ (\bibinfo
  {institution} {{LIGO Document T0900288-v3}},\ \bibinfo {year}
  {2009})\BibitemShut {NoStop}%
\bibitem [{\citenamefont {Cutler}\ and\ \citenamefont
  {Flanagan}(1994)}]{Cutler:1994ys}%
  \BibitemOpen
  \bibfield  {author} {\bibinfo {author} {\bibfnamefont {C.}~\bibnamefont
  {Cutler}}\ and\ \bibinfo {author} {\bibfnamefont {E.~E.}\ \bibnamefont
  {Flanagan}},\ }\href {\doibase 10.1103/PhysRevD.49.2658} {\bibfield
  {journal} {\bibinfo  {journal} {Phys.Rev.}\ }\textbf {\bibinfo {volume}
  {D49}},\ \bibinfo {pages} {2658} (\bibinfo {year} {1994})},\ \Eprint
  {http://arxiv.org/abs/gr-qc/9402014} {arXiv:gr-qc/9402014 [gr-qc]}
  \BibitemShut {NoStop}%
\bibitem [{\citenamefont {Droz}\ \emph {et~al.}(1999)\citenamefont {Droz},
  \citenamefont {Knapp}, \citenamefont {Poisson},\ and\ \citenamefont
  {Owen}}]{Droz:1999qx}%
  \BibitemOpen
  \bibfield  {author} {\bibinfo {author} {\bibfnamefont {S.}~\bibnamefont
  {Droz}}, \bibinfo {author} {\bibfnamefont {D.~J.}\ \bibnamefont {Knapp}},
  \bibinfo {author} {\bibfnamefont {E.}~\bibnamefont {Poisson}}, \ and\
  \bibinfo {author} {\bibfnamefont {B.~J.}\ \bibnamefont {Owen}},\ }\href
  {\doibase 10.1103/PhysRevD.59.124016} {\bibfield  {journal} {\bibinfo
  {journal} {Phys.Rev.}\ }\textbf {\bibinfo {volume} {D59}},\ \bibinfo {pages}
  {124016} (\bibinfo {year} {1999})},\ \Eprint
  {http://arxiv.org/abs/gr-qc/9901076} {arXiv:gr-qc/9901076 [gr-qc]}
  \BibitemShut {NoStop}%
\bibitem [{\citenamefont {Blanchet}\ \emph
  {et~al.}(2004{\natexlab{a}})\citenamefont {Blanchet}, \citenamefont {Damour},
  \citenamefont {Esposito-Farese},\ and\ \citenamefont
  {Iyer}}]{PNFluxEnergy3PN01}%
  \BibitemOpen
  \bibfield  {author} {\bibinfo {author} {\bibfnamefont {L.}~\bibnamefont
  {Blanchet}}, \bibinfo {author} {\bibfnamefont {T.}~\bibnamefont {Damour}},
  \bibinfo {author} {\bibfnamefont {G.}~\bibnamefont {Esposito-Farese}}, \ and\
  \bibinfo {author} {\bibfnamefont {B.~R.}\ \bibnamefont {Iyer}},\ }\href
  {\doibase 10.1103/PhysRevLett.93.091101} {\bibfield  {journal} {\bibinfo
  {journal} {Phys.Rev.Lett.}\ }\textbf {\bibinfo {volume} {93}},\ \bibinfo
  {pages} {091101} (\bibinfo {year} {2004}{\natexlab{a}})},\ \Eprint
  {http://arxiv.org/abs/gr-qc/0406012} {arXiv:gr-qc/0406012 [gr-qc]}
  \BibitemShut {NoStop}%
\bibitem [{\citenamefont {Blanchet}\ and\ \citenamefont
  {Iyer}(2005)}]{PNFluxEnergy3PN02}%
  \BibitemOpen
  \bibfield  {author} {\bibinfo {author} {\bibfnamefont {L.}~\bibnamefont
  {Blanchet}}\ and\ \bibinfo {author} {\bibfnamefont {B.~R.}\ \bibnamefont
  {Iyer}},\ }\href {\doibase 10.1103/PhysRevD.71.024004} {\bibfield  {journal}
  {\bibinfo  {journal} {Phys.Rev.}\ }\textbf {\bibinfo {volume} {D71}},\
  \bibinfo {pages} {024004} (\bibinfo {year} {2005})},\ \Eprint
  {http://arxiv.org/abs/gr-qc/0409094} {arXiv:gr-qc/0409094 [gr-qc]}
  \BibitemShut {NoStop}%
\bibitem [{\citenamefont {Jaranowski}\ and\ \citenamefont
  {Schaefer}(2000)}]{Jaranowski:1999qd}%
  \BibitemOpen
  \bibfield  {author} {\bibinfo {author} {\bibfnamefont {P.}~\bibnamefont
  {Jaranowski}}\ and\ \bibinfo {author} {\bibfnamefont {G.}~\bibnamefont
  {Schaefer}},\ }\href@noop {} {\bibfield  {journal} {\bibinfo  {journal}
  {Annalen Phys.}\ }\textbf {\bibinfo {volume} {9}},\ \bibinfo {pages} {378}
  (\bibinfo {year} {2000})},\ \Eprint {http://arxiv.org/abs/gr-qc/0003054}
  {arXiv:gr-qc/0003054 [gr-qc]} \BibitemShut {NoStop}%
\bibitem [{\citenamefont {Jaranowski}\ and\ \citenamefont
  {Schafer}(1999)}]{Jaranowski:1999ye}%
  \BibitemOpen
  \bibfield  {author} {\bibinfo {author} {\bibfnamefont {P.}~\bibnamefont
  {Jaranowski}}\ and\ \bibinfo {author} {\bibfnamefont {G.}~\bibnamefont
  {Schafer}},\ }\href {\doibase 10.1103/PhysRevD.60.124003} {\bibfield
  {journal} {\bibinfo  {journal} {Phys.Rev.}\ }\textbf {\bibinfo {volume}
  {D60}},\ \bibinfo {pages} {124003} (\bibinfo {year} {1999})},\ \Eprint
  {http://arxiv.org/abs/gr-qc/9906092} {arXiv:gr-qc/9906092 [gr-qc]}
  \BibitemShut {NoStop}%
\bibitem [{\citenamefont {Damour}\ \emph {et~al.}(2001)\citenamefont {Damour},
  \citenamefont {Jaranowski},\ and\ \citenamefont {Schaefer}}]{Damour:2001bu}%
  \BibitemOpen
  \bibfield  {author} {\bibinfo {author} {\bibfnamefont {T.}~\bibnamefont
  {Damour}}, \bibinfo {author} {\bibfnamefont {P.}~\bibnamefont {Jaranowski}},
  \ and\ \bibinfo {author} {\bibfnamefont {G.}~\bibnamefont {Schaefer}},\
  }\href {\doibase 10.1016/S0370-2693(01)00642-6} {\bibfield  {journal}
  {\bibinfo  {journal} {Phys.Lett.}\ }\textbf {\bibinfo {volume} {B513}},\
  \bibinfo {pages} {147} (\bibinfo {year} {2001})},\ \Eprint
  {http://arxiv.org/abs/gr-qc/0105038} {arXiv:gr-qc/0105038 [gr-qc]}
  \BibitemShut {NoStop}%
\bibitem [{\citenamefont {Kidder}(2008)}]{KidderPN}%
  \BibitemOpen
  \bibfield  {author} {\bibinfo {author} {\bibfnamefont {L.~E.}\ \bibnamefont
  {Kidder}},\ }\href {\doibase 10.1103/PhysRevD.77.044016} {\bibfield
  {journal} {\bibinfo  {journal} {Phys.Rev.}\ }\textbf {\bibinfo {volume}
  {D77}},\ \bibinfo {pages} {044016} (\bibinfo {year} {2008})},\ \Eprint
  {http://arxiv.org/abs/0710.0614} {arXiv:0710.0614 [gr-qc]} \BibitemShut
  {NoStop}%
\bibitem [{\citenamefont {Blanchet}\ \emph {et~al.}(2008)\citenamefont
  {Blanchet}, \citenamefont {Faye}, \citenamefont {Iyer},\ and\ \citenamefont
  {Sinha}}]{Blanchet3PN}%
  \BibitemOpen
  \bibfield  {author} {\bibinfo {author} {\bibfnamefont {L.}~\bibnamefont
  {Blanchet}}, \bibinfo {author} {\bibfnamefont {G.}~\bibnamefont {Faye}},
  \bibinfo {author} {\bibfnamefont {B.~R.}\ \bibnamefont {Iyer}}, \ and\
  \bibinfo {author} {\bibfnamefont {S.}~\bibnamefont {Sinha}},\ }\href
  {\doibase 10.1088/0264-9381/25/16/165003} {\bibfield  {journal} {\bibinfo
  {journal} {Class.Quant.Grav.}\ }\textbf {\bibinfo {volume} {25}},\ \bibinfo
  {pages} {165003} (\bibinfo {year} {2008})},\ \bibinfo {note} {57 pages, no
  figures},\ \Eprint {http://arxiv.org/abs/0802.1249} {arXiv:0802.1249 [gr-qc]}
  \BibitemShut {NoStop}%
\bibitem [{\citenamefont {Buonanno}\ \emph
  {et~al.}(2009{\natexlab{b}})\citenamefont {Buonanno}, \citenamefont {Iyer},
  \citenamefont {Ochsner}, \citenamefont {Pan},\ and\ \citenamefont
  {Sathyaprakash}}]{CompTemplates2009}%
  \BibitemOpen
  \bibfield  {author} {\bibinfo {author} {\bibfnamefont {A.}~\bibnamefont
  {Buonanno}}, \bibinfo {author} {\bibfnamefont {B.~R.}~\bibnamefont {Iyer}},
  \bibinfo {author} {\bibfnamefont {E.}~\bibnamefont {Ochsner}}, \bibinfo
  {author} {\bibfnamefont {Y.}~\bibnamefont {Pan}}, \ and\ \bibinfo {author}
  {\bibfnamefont {B.~S.}~\bibnamefont {Sathyaprakash}},\ }\href {\doibase
  10.1103/PhysRevD.80.084043} {\bibfield  {journal} {\bibinfo  {journal}
  {Phys.Rev.}\ }\textbf {\bibinfo {volume} {D80}},\ \bibinfo {pages} {084043}
  (\bibinfo {year} {2009}{\natexlab{b}})},\ \Eprint
  {http://arxiv.org/abs/0907.0700} {arXiv:0907.0700 [gr-qc]} \BibitemShut
  {NoStop}%
\bibitem [{\citenamefont {McKechan}(2011)}]{McKechan:2011ps}%
  \BibitemOpen
  \bibfield  {author} {\bibinfo {author} {\bibfnamefont {D.}~\bibnamefont
  {McKechan}},\ }\href@noop {} {\  (\bibinfo {year} {2011})},\ \bibinfo {note}
  {ph.D. thesis},\ \Eprint {http://arxiv.org/abs/1102.1749} {arXiv:1102.1749
  [gr-qc]} \BibitemShut {NoStop}%
\bibitem [{\citenamefont {Pekowsky}\ \emph {et~al.}(2012)\citenamefont
  {Pekowsky}, \citenamefont {Healy}, \citenamefont {Shoemaker},\ and\
  \citenamefont {Laguna}}]{Pekowsky:2012sr}%
  \BibitemOpen
  \bibfield  {author} {\bibinfo {author} {\bibfnamefont {L.}~\bibnamefont
  {Pekowsky}}, \bibinfo {author} {\bibfnamefont {J.}~\bibnamefont {Healy}},
  \bibinfo {author} {\bibfnamefont {D.}~\bibnamefont {Shoemaker}}, \ and\
  \bibinfo {author} {\bibfnamefont {P.}~\bibnamefont {Laguna}},\ }\href@noop {}
  {\  (\bibinfo {year} {2012})},\ \Eprint {http://arxiv.org/abs/1210.1891}
  {arXiv:1210.1891 [gr-qc]} \BibitemShut {NoStop}%
\bibitem [{\citenamefont {Blanchet}(2006)}]{PNtheoryLivingReviewBlanchet}%
  \BibitemOpen
  \bibfield  {author} {\bibinfo {author} {\bibfnamefont {L.}~\bibnamefont
  {Blanchet}},\ }\href@noop {} {\bibfield  {journal} {\bibinfo  {journal}
  {Living Rev.Rel.}\ }\textbf {\bibinfo {volume} {9}},\ \bibinfo {pages} {4}
  (\bibinfo {year} {2006})}\BibitemShut {NoStop}%
\bibitem [{\citenamefont {Pretorius}(2005)}]{Pretorius2005}%
  \BibitemOpen
  \bibfield  {author} {\bibinfo {author} {\bibfnamefont {F.}~\bibnamefont
  {Pretorius}},\ }\href {\doibase 10.1103/PhysRevLett.95.121101} {\bibfield
  {journal} {\bibinfo  {journal} {Phys.Rev.Lett.}\ }\textbf {\bibinfo {volume}
  {95}},\ \bibinfo {pages} {121101} (\bibinfo {year} {2005})},\ \Eprint
  {http://arxiv.org/abs/gr-qc/0507014} {arXiv:gr-qc/0507014 [gr-qc]}
  \BibitemShut {NoStop}%
\bibitem [{\citenamefont {Pretorius}(2006)}]{Pretorius2006}%
  \BibitemOpen
  \bibfield  {author} {\bibinfo {author} {\bibfnamefont {F.}~\bibnamefont
  {Pretorius}},\ }\href {\doibase 10.1088/0264-9381/23/16/S13} {\bibfield
  {journal} {\bibinfo  {journal} {Class.Quant.Grav.}\ }\textbf {\bibinfo
  {volume} {23}},\ \bibinfo {pages} {S529} (\bibinfo {year} {2006})},\ \Eprint
  {http://arxiv.org/abs/gr-qc/0602115} {arXiv:gr-qc/0602115 [gr-qc]}
  \BibitemShut {NoStop}%
\bibitem [{\citenamefont {Scheel}\ \emph {et~al.}(2009)\citenamefont {Scheel},
  \citenamefont {Boyle}, \citenamefont {Chu}, \citenamefont {Kidder},
  \citenamefont {Matthews} \emph {et~al.}}]{BBHNRScheel}%
  \BibitemOpen
  \bibfield  {author} {\bibinfo {author} {\bibfnamefont {M.~A.}\ \bibnamefont
  {Scheel}}, \bibinfo {author} {\bibfnamefont {M.}~\bibnamefont {Boyle}},
  \bibinfo {author} {\bibfnamefont {T.}~\bibnamefont {Chu}}, \bibinfo {author}
  {\bibfnamefont {L.~E.}\ \bibnamefont {Kidder}}, \bibinfo {author}
  {\bibfnamefont {K.~D.}\ \bibnamefont {Matthews}},  \emph {et~al.},\ }\href
  {\doibase 10.1103/PhysRevD.79.024003} {\bibfield  {journal} {\bibinfo
  {journal} {Phys.Rev.}\ }\textbf {\bibinfo {volume} {D79}},\ \bibinfo {pages}
  {024003} (\bibinfo {year} {2009})},\ \Eprint {http://arxiv.org/abs/0810.1767}
  {arXiv:0810.1767 [gr-qc]} \BibitemShut {NoStop}%
\bibitem [{\citenamefont {Gonzalez}\ \emph {et~al.}(2009)\citenamefont
  {Gonzalez}, \citenamefont {Sperhake},\ and\ \citenamefont
  {Brugmann}}]{BBHNRGonzalezq10}%
  \BibitemOpen
  \bibfield  {author} {\bibinfo {author} {\bibfnamefont {J.~A.}\ \bibnamefont
  {Gonzalez}}, \bibinfo {author} {\bibfnamefont {U.}~\bibnamefont {Sperhake}},
  \ and\ \bibinfo {author} {\bibfnamefont {B.}~\bibnamefont {Brugmann}},\
  }\href {\doibase 10.1103/PhysRevD.79.124006} {\bibfield  {journal} {\bibinfo
  {journal} {Phys.Rev.}\ }\textbf {\bibinfo {volume} {D79}},\ \bibinfo {pages}
  {124006} (\bibinfo {year} {2009})},\ \Eprint {http://arxiv.org/abs/0811.3952}
  {arXiv:0811.3952 [gr-qc]} \BibitemShut {NoStop}%
\bibitem [{\citenamefont {Pollney}\ \emph {et~al.}(2011)\citenamefont
  {Pollney}, \citenamefont {Reisswig}, \citenamefont {Schnetter}, \citenamefont
  {Dorband},\ and\ \citenamefont {Diener}}]{BBHNRPollney}%
  \BibitemOpen
  \bibfield  {author} {\bibinfo {author} {\bibfnamefont {D.}~\bibnamefont
  {Pollney}}, \bibinfo {author} {\bibfnamefont {C.}~\bibnamefont {Reisswig}},
  \bibinfo {author} {\bibfnamefont {E.}~\bibnamefont {Schnetter}}, \bibinfo
  {author} {\bibfnamefont {N.}~\bibnamefont {Dorband}}, \ and\ \bibinfo
  {author} {\bibfnamefont {P.}~\bibnamefont {Diener}},\ }\href {\doibase
  10.1103/PhysRevD.83.044045} {\bibfield  {journal} {\bibinfo  {journal}
  {Phys.Rev.}\ }\textbf {\bibinfo {volume} {D83}},\ \bibinfo {pages} {044045}
  (\bibinfo {year} {2011})},\ \Eprint {http://arxiv.org/abs/0910.3803}
  {arXiv:0910.3803 [gr-qc]} \BibitemShut {NoStop}%
\bibitem [{\citenamefont {Lousto}\ \emph {et~al.}(2010)\citenamefont {Lousto},
  \citenamefont {Nakano}, \citenamefont {Zlochower},\ and\ \citenamefont
  {Campanelli}}]{BBHNRLoustoq10}%
  \BibitemOpen
  \bibfield  {author} {\bibinfo {author} {\bibfnamefont {C.~O.}\ \bibnamefont
  {Lousto}}, \bibinfo {author} {\bibfnamefont {H.}~\bibnamefont {Nakano}},
  \bibinfo {author} {\bibfnamefont {Y.}~\bibnamefont {Zlochower}}, \ and\
  \bibinfo {author} {\bibfnamefont {M.}~\bibnamefont {Campanelli}},\ }\href
  {\doibase 10.1103/PhysRevLett.104.211101} {\bibfield  {journal} {\bibinfo
  {journal} {Phys.Rev.Lett.}\ }\textbf {\bibinfo {volume} {104}},\ \bibinfo
  {pages} {211101} (\bibinfo {year} {2010})},\ \Eprint
  {http://arxiv.org/abs/1001.2316} {arXiv:1001.2316 [gr-qc]} \BibitemShut
  {NoStop}%
\bibitem [{\citenamefont {Buchman}\ \emph {et~al.}(2012)\citenamefont
  {Buchman}, \citenamefont {Pfeiffer}, \citenamefont {Scheel},\ and\
  \citenamefont {Szilagyi}}]{Buchman:2012dw}%
  \BibitemOpen
  \bibfield  {author} {\bibinfo {author} {\bibfnamefont {L.~T.}\ \bibnamefont
  {Buchman}}, \bibinfo {author} {\bibfnamefont {H.~P.}\ \bibnamefont
  {Pfeiffer}}, \bibinfo {author} {\bibfnamefont {M.~A.}\ \bibnamefont
  {Scheel}}, \ and\ \bibinfo {author} {\bibfnamefont {B.}~\bibnamefont
  {Szilagyi}},\ }\href@noop {} {\  (\bibinfo {year} {2012})},\ \Eprint
  {http://arxiv.org/abs/1206.3015} {arXiv:1206.3015 [gr-qc]} \BibitemShut
  {NoStop}%
\bibitem [{\citenamefont {Ajith}\ \emph {et~al.}(2012)\citenamefont {Ajith},
  \citenamefont {Boyle}, \citenamefont {Brown}, \citenamefont {Brugmann},
  \citenamefont {Buchman} \emph {et~al.}}]{Ajith:2012tt}%
  \BibitemOpen
  \bibfield  {author} {\bibinfo {author} {\bibfnamefont {P.}~\bibnamefont
  {Ajith}}, \bibinfo {author} {\bibfnamefont {M.}~\bibnamefont {Boyle}},
  \bibinfo {author} {\bibfnamefont {D.~A.}\ \bibnamefont {Brown}}, \bibinfo
  {author} {\bibfnamefont {B.}~\bibnamefont {Brugmann}}, \bibinfo {author}
  {\bibfnamefont {L.~T.}\ \bibnamefont {Buchman}},  \emph {et~al.},\ }\href
  {\doibase 10.1088/0264-9381/29/12/124001} {\bibfield  {journal} {\bibinfo
  {journal} {Class.Quant.Grav.}\ }\textbf {\bibinfo {volume} {29}},\ \bibinfo
  {pages} {124001} (\bibinfo {year} {2012})},\ \Eprint
  {http://arxiv.org/abs/1201.5319} {arXiv:1201.5319 [gr-qc]} \BibitemShut
  {NoStop}%
\bibitem [{\citenamefont {Damour}\ \emph {et~al.}(2000)\citenamefont {Damour},
  \citenamefont {Jaranowski},\ and\ \citenamefont {Schafer}}]{PadeAD}%
  \BibitemOpen
  \bibfield  {author} {\bibinfo {author} {\bibfnamefont {T.}~\bibnamefont
  {Damour}}, \bibinfo {author} {\bibfnamefont {P.}~\bibnamefont {Jaranowski}},
  \ and\ \bibinfo {author} {\bibfnamefont {G.}~\bibnamefont {Schafer}},\
  }\href {\doibase 10.1103/PhysRevD.62.084011} {\bibfield  {journal} {\bibinfo
  {journal} {Phys.Rev.}\ }\textbf {\bibinfo {volume} {D62}},\ \bibinfo {pages}
  {084011} (\bibinfo {year} {2000})},\ \Eprint
  {http://arxiv.org/abs/gr-qc/0005034} {arXiv:gr-qc/0005034 [gr-qc]}
  \BibitemShut {NoStop}%
\bibitem [{\citenamefont {Damour}\ \emph {et~al.}(2009)\citenamefont {Damour},
  \citenamefont {Iyer},\ and\ \citenamefont {Nagar}}]{DamourFluxhlm01}%
  \BibitemOpen
  \bibfield  {author} {\bibinfo {author} {\bibfnamefont {T.}~\bibnamefont
  {Damour}}, \bibinfo {author} {\bibfnamefont {B.~R.}\ \bibnamefont {Iyer}}, \
  and\ \bibinfo {author} {\bibfnamefont {A.}~\bibnamefont {Nagar}},\ }\href
  {\doibase 10.1103/PhysRevD.79.064004} {\bibfield  {journal} {\bibinfo
  {journal} {Phys.Rev.}\ }\textbf {\bibinfo {volume} {D79}},\ \bibinfo {pages}
  {064004} (\bibinfo {year} {2009})},\ \Eprint {http://arxiv.org/abs/0811.2069}
  {arXiv:0811.2069 [gr-qc]} \BibitemShut {NoStop}%
\bibitem [{\citenamefont {Buonanno}\ and\ \citenamefont
  {Damour}(1999{\natexlab{b}})}]{BuonannoEOBTerms}%
  \BibitemOpen
  \bibfield  {author} {\bibinfo {author} {\bibfnamefont {A.}~\bibnamefont
  {Buonanno}}\ and\ \bibinfo {author} {\bibfnamefont {T.}~\bibnamefont
  {Damour}},\ }\href {\doibase 10.1103/PhysRevD.59.084006} {\bibfield
  {journal} {\bibinfo  {journal} {Phys.Rev.}\ }\textbf {\bibinfo {volume}
  {D59}},\ \bibinfo {pages} {084006} (\bibinfo {year} {1999}{\natexlab{b}})},\
  \Eprint {http://arxiv.org/abs/gr-qc/9811091} {arXiv:gr-qc/9811091 [gr-qc]}
  \BibitemShut {NoStop}%
\bibitem [{lal()}]{lal}%
  \BibitemOpen
  \href@noop {} {\enquote {\bibinfo {title} {L{SC} {A}lgorithm {L}ibrary},}\
  }\bibinfo {howpublished}
  {\url{https://www.lsc-group.phys.uwm.edu/daswg/projects/lalsuite.html}}\BibitemShut
  {NoStop}%
\bibitem [{\citenamefont {Mathews}\ and\ \citenamefont
  {Walker}(1970)}]{MatthewsWalker}%
  \BibitemOpen
  \bibfield  {author} {\bibinfo {author} {\bibfnamefont {J.}~\bibnamefont
  {Mathews}}\ and\ \bibinfo {author} {\bibfnamefont {R.~L.}\ \bibnamefont
  {Walker}},\ }\href@noop {} {\emph {\bibinfo {title} {Mathematical Methods of
  Physics}}}\ (\bibinfo  {publisher} {W.A. Benjamin; 2nd edition},\ \bibinfo
  {year} {1970})\BibitemShut {NoStop}%
\bibitem [{\citenamefont {Blanchet}\ \emph {et~al.}(1995)\citenamefont
  {Blanchet}, \citenamefont {Damour}, \citenamefont {Iyer}, \citenamefont
  {Will},\ and\ \citenamefont {Wiseman}}]{GW2PN}%
  \BibitemOpen
  \bibfield  {author} {\bibinfo {author} {\bibfnamefont {L.}~\bibnamefont
  {Blanchet}}, \bibinfo {author} {\bibfnamefont {T.}~\bibnamefont {Damour}},
  \bibinfo {author} {\bibfnamefont {B.~R.}\ \bibnamefont {Iyer}}, \bibinfo
  {author} {\bibfnamefont {C.~M.}\ \bibnamefont {Will}}, \ and\ \bibinfo
  {author} {\bibfnamefont {A.~G.}~\bibnamefont {Wiseman}},\ }\href {\doibase
  10.1103/PhysRevLett.74.3515} {\bibfield  {journal} {\bibinfo  {journal}
  {Phys.Rev.Lett.}\ }\textbf {\bibinfo {volume} {74}},\ \bibinfo {pages} {3515}
  (\bibinfo {year} {1995})},\ \Eprint {http://arxiv.org/abs/gr-qc/9501027}
  {arXiv:gr-qc/9501027 [gr-qc]} \BibitemShut {NoStop}%
\bibitem [{\citenamefont {Blanchet}\ \emph {et~al.}(2002)\citenamefont
  {Blanchet}, \citenamefont {Faye}, \citenamefont {Iyer},\ and\ \citenamefont
  {Joguet}}]{Blanchet:2001ax}%
  \BibitemOpen
  \bibfield  {author} {\bibinfo {author} {\bibfnamefont {L.}~\bibnamefont
  {Blanchet}}, \bibinfo {author} {\bibfnamefont {G.}~\bibnamefont {Faye}},
  \bibinfo {author} {\bibfnamefont {B.~R.}\ \bibnamefont {Iyer}}, \ and\
  \bibinfo {author} {\bibfnamefont {B.}~\bibnamefont {Joguet}},\ }\href
  {\doibase 10.1103/PhysRevD.71.129902, 10.1103/PhysRevD.65.061501} {\bibfield
  {journal} {\bibinfo  {journal} {Phys.Rev.}\ }\textbf {\bibinfo {volume}
  {D65}},\ \bibinfo {pages} {061501} (\bibinfo {year} {2002})},\ \Eprint
  {http://arxiv.org/abs/gr-qc/0105099} {arXiv:gr-qc/0105099 [gr-qc]}
  \BibitemShut {NoStop}%
\bibitem [{\citenamefont {Blanchet}\ \emph
  {et~al.}(2004{\natexlab{b}})\citenamefont {Blanchet}, \citenamefont {Damour},
  \citenamefont {Esposito-Farese},\ and\ \citenamefont
  {Iyer}}]{Blanchet:2004ek}%
  \BibitemOpen
  \bibfield  {author} {\bibinfo {author} {\bibfnamefont {L.}~\bibnamefont
  {Blanchet}}, \bibinfo {author} {\bibfnamefont {T.}~\bibnamefont {Damour}},
  \bibinfo {author} {\bibfnamefont {G.}~\bibnamefont {Esposito-Farese}}, \ and\
  \bibinfo {author} {\bibfnamefont {B.~R.}\ \bibnamefont {Iyer}},\ }\href
  {\doibase 10.1103/PhysRevLett.93.091101} {\bibfield  {journal} {\bibinfo
  {journal} {Phys.Rev.Lett.}\ }\textbf {\bibinfo {volume} {93}},\ \bibinfo
  {pages} {091101} (\bibinfo {year} {2004}{\natexlab{b}})},\ \Eprint
  {http://arxiv.org/abs/gr-qc/0406012} {arXiv:gr-qc/0406012 [gr-qc]}
  \BibitemShut {NoStop}%
\bibitem [{\citenamefont {Poisson}\ and\ \citenamefont
  {Will}(1995)}]{Poisson:1995ef}%
  \BibitemOpen
  \bibfield  {author} {\bibinfo {author} {\bibfnamefont {E.}~\bibnamefont
  {Poisson}}\ and\ \bibinfo {author} {\bibfnamefont {C.~M.}\ \bibnamefont
  {Will}},\ }\href {\doibase 10.1103/PhysRevD.52.848} {\bibfield  {journal}
  {\bibinfo  {journal} {Phys.Rev.}\ }\textbf {\bibinfo {volume} {D52}},\
  \bibinfo {pages} {848} (\bibinfo {year} {1995})},\ \Eprint
  {http://arxiv.org/abs/gr-qc/9502040} {arXiv:gr-qc/9502040 [gr-qc]}
  \BibitemShut {NoStop}%
\bibitem [{\citenamefont {Lindblom}\ \emph {et~al.}(2008)\citenamefont
  {Lindblom}, \citenamefont {Owen},\ and\ \citenamefont
  {Brown}}]{WaveformAccuracy2008}%
  \BibitemOpen
  \bibfield  {author} {\bibinfo {author} {\bibfnamefont {L.}~\bibnamefont
  {Lindblom}}, \bibinfo {author} {\bibfnamefont {B.~J.}\ \bibnamefont {Owen}},
  \ and\ \bibinfo {author} {\bibfnamefont {D.~A.}\ \bibnamefont {Brown}},\
  }\href {\doibase 10.1103/PhysRevD.78.124020} {\bibfield  {journal} {\bibinfo
  {journal} {Phys.Rev.}\ }\textbf {\bibinfo {volume} {D78}},\ \bibinfo {pages}
  {124020} (\bibinfo {year} {2008})},\ \Eprint {http://arxiv.org/abs/0809.3844}
  {arXiv:0809.3844 [gr-qc]} \BibitemShut {NoStop}%
\bibitem [{\citenamefont {Lindblom}\ \emph {et~al.}(2010)\citenamefont
  {Lindblom}, \citenamefont {Baker},\ and\ \citenamefont
  {Owen}}]{WaveformAccuracy2010}%
  \BibitemOpen
  \bibfield  {author} {\bibinfo {author} {\bibfnamefont {L.}~\bibnamefont
  {Lindblom}}, \bibinfo {author} {\bibfnamefont {J.~G.}\ \bibnamefont {Baker}},
  \ and\ \bibinfo {author} {\bibfnamefont {B.~J.}\ \bibnamefont {Owen}},\
  }\href {\doibase 10.1103/PhysRevD.82.084020} {\bibfield  {journal} {\bibinfo
  {journal} {Phys.Rev.}\ }\textbf {\bibinfo {volume} {D82}},\ \bibinfo {pages}
  {084020} (\bibinfo {year} {2010})},\ \Eprint {http://arxiv.org/abs/1008.1803}
  {arXiv:1008.1803 [gr-qc]} \BibitemShut {NoStop}%
\bibitem [{\citenamefont {Buonanno}\ \emph {et~al.}(2007)\citenamefont
  {Buonanno}, \citenamefont {Pan}, \citenamefont {Baker}, \citenamefont
  {Centrella}, \citenamefont {Kelly} \emph {et~al.}}]{Buonanno:2007pf}%
  \BibitemOpen
  \bibfield  {author} {\bibinfo {author} {\bibfnamefont {A.}~\bibnamefont
  {Buonanno}}, \bibinfo {author} {\bibfnamefont {Y.}~\bibnamefont {Pan}},
  \bibinfo {author} {\bibfnamefont {J.~G.}\ \bibnamefont {Baker}}, \bibinfo
  {author} {\bibfnamefont {J.}~\bibnamefont {Centrella}}, \bibinfo {author}
  {\bibfnamefont {B.~J.}\ \bibnamefont {Kelly}},  \emph {et~al.},\ }\href
  {\doibase 10.1103/PhysRevD.76.104049} {\bibfield  {journal} {\bibinfo
  {journal} {Phys.Rev.}\ }\textbf {\bibinfo {volume} {D76}},\ \bibinfo {pages}
  {104049} (\bibinfo {year} {2007})},\ \Eprint {http://arxiv.org/abs/0706.3732}
  {arXiv:0706.3732 [gr-qc]} \BibitemShut {NoStop}%
\bibitem [{\citenamefont {Sathyaprakash}\ and\ \citenamefont
  {Schutz}(2009)}]{Sathyaprakash:2009xs}%
  \BibitemOpen
  \bibfield  {author} {\bibinfo {author} {\bibfnamefont {B.}~\bibnamefont
  {Sathyaprakash}}\ and\ \bibinfo {author} {\bibfnamefont {B.}~\bibnamefont
  {Schutz}},\ }\href@noop {} {\bibfield  {journal} {\bibinfo  {journal} {Living
  Rev.Rel.}\ }\textbf {\bibinfo {volume} {12}},\ \bibinfo {pages} {2} (\bibinfo
  {year} {2009})},\ \Eprint {http://arxiv.org/abs/0903.0338} {arXiv:0903.0338
  [gr-qc]} \BibitemShut {NoStop}%
\bibitem{Finn:1992xs} 
  L.~S.~Finn and D.~F.~Chernoff,
  Phys.\ Rev.\ D {\bf 47}, 2198 (1993)
  [gr-qc/9301003].
\bibitem{Buonanno:2005xu} 
  A.~Buonanno, Y.~Chen and T.~Damour,
  Phys.\ Rev.\ D {\bf 74}, 104005 (2006)
  [gr-qc/0508067].
\end{thebibliography}

%

\end{document}